\documentclass[12pt]{iopart}

\usepackage{iopams}
\usepackage{amstext}
\usepackage{mathabx}
\usepackage{graphicx}
\usepackage{verbatim} 
\usepackage{color}

\newcommand{\avg}[1]{\mbox{$\langle#1\rangle$}}

\def\be{\begin{equation}}
\def\ee{\end{equation}}
\def\bea{\begin{eqnarray}}
\def\eea{\end{eqnarray}}
\newcommand{\opdagger}[2]{\mbox{$\hat{#1}_{#2}^{\dagger}$}}
\newcommand{\op}[2]{\mbox{$\hat{#1}_{#2}$}}

\newcommand{\nbar}{\avg{\op{n}{}}}

\def\be{\begin{equation}}
\def\ee{\end{equation}}
\def\bea{\begin{eqnarray}}
\def\eea{\end{eqnarray}}

\renewcommand{\avg}[1]{\mbox{$\langle#1\rangle$}}

\renewcommand{\opdagger}[2]{\mbox{$\hat{#1}_{#2}^{\dagger}$}}
\renewcommand{\op}[2]{\mbox{$\hat{#1}_{#2}$}}

\usepackage{hhline}
\usepackage{threeparttable}
\usepackage{supertabular}
\usepackage{multirow}
\usepackage{tabularx}

\newcolumntype{Y}{>{\centering\arraybackslash}X}

\begin{document}

\pagenumbering{arabic}

\title{Laser noise in cavity-optomechanical cooling and thermometry}

\author{Amir H. Safavi-Naeini$^1$, Jasper Chan$^1$, Jeff T. Hill$^1$,  Simon Gr\"oblacher$^{1,2}$, Haixing Miao$^3$, Yanbei Chen$^3$, Markus Aspelmeyer$^4$, Oskar Painter$^1$}

\address{$^1$Kavli Nanoscience Institute and Thomas J. Watson, Sr., Laboratory of Applied Physics, California Institute of Technology, Pasadena CA 91125, USA}
\address{$^2$Institute for Quantum Information and Matter, California Institute of Technology, 1200 E. California Blvd., Pasadena, CA 91125, USA}
\address{$^3$Theoretical Astrophysics 350-17, California Institute of Technology, Pasadena, CA 91125, USA}
\address{$^4$Vienna Center for Quantum Science and Technology (VCQ), Faculty of Physics, University of Vienna, Boltzmanngasse 5, A-1090 Vienna, Austria}
\ead{safavi@caltech.edu, opainter@caltech.edu}

\date{\today}
%

\begin{abstract}
We review and study the roles of quantum and classical fluctuations in recent cavity-optomechanical experiments which have now reached the quantum regime (mechanical phonon occupancy $\lesssim 1$) using resolved sideband laser cooling. In particular, both the laser noise heating of the mechanical resonator and the form of the optically transduced mechanical spectra, modified by quantum and classical laser noise squashing, are derived under various measurement conditions.  Using this theory, we analyze recent ground-state laser cooling and motional sideband asymmetry experiments with nanoscale optomechanical crystal resonators.  
\end{abstract}

\maketitle

\section{Introduction}

Mechanical resonators provide an interesting and useful system for the study of quantum theory in the mesoscopic and macroscopic scales~\cite{Braginsky1977,Caves1980a,Caves1980b,Braginsky1995,Clerk2008,Clerk2010}. Technological progress in nano- and microfabrication have made coupling of motion to a wide variety of systems possible, and experiments demonstrating mechanical resonators coupled to optical cavities~\cite{Arcizet2006b,Gigan2006,Kleckner2006a,Schliesser2006,Groeblacher2009a,Eichenfield2009a,Eichenfield2009b}, microwave resonators~\cite{Regal2008,Rocheleau2010,Teufel2011b,Massel2011a}, superconducting qubits~\cite{LaHaye2009,OConnell2010,Pirkkalainen2012}, cold atoms~\cite{Camerer2011a}, and defect centers in diamond~\cite{Arcizet2011} have been forthcoming. These advances have led to the cooling of mechanical systems to their quantum ground states~\cite{OConnell2010,Teufel2011b,Chan2011}, and the observation of nonclassical behaviour~\cite{OConnell2010,Safavi-Naeini2012,Brahms2012}. Of all the systems to which mechanical systems have been shown to couple, it has long been realized~\cite{Braginsky1977} that light has many distinct advantages. Most importantly light sources can be made quantum limited (lasers) and lack thermal noise even at room temperature. This allows, for example, to cool optomechanical systems close to their ground states using radiation pressure forces, and recently experiments starting with modest cryogenic pre-cooling~\cite{Groeblacher2009a,Chan2011,Safavi-Naeini2012,Verhagen2012}, have been successful at achieving this task.

Considering the importance of cooling and thermometry in cavity-optomechanical experiments, it is important to understand the noise processes which can lead to both heating and systematic errors in thermometry. In this paper we focus on analyzing these effects theoretically, and compare to recently performed experiments with nanoscale optomechanical crystal devices. In Section~\ref{sec:theory} a theoretical treatment of optomechanical cooling and thermometry particularly suited for understanding the propagation of noise is given. The regime of operation analyzed is that relevant to our recent experiments~\cite{Chan2011,Safavi-Naeini2012}, i.e. the driven weak-coupling, sideband-resolved regime where the cavity decay rate is larger than all other rates in the system except for the mechanical frequency. In this regime, the standard input-ouput formalism~\cite{Gardiner1985,Collett1984,Gardiner2004} for analyzing the linearized system is applied. Two different methods of thermometry used in recent experiments~\cite{Chan2011,Safavi-Naeini2012} are treated in sections \ref{ss:simple_therm} and \ref{ss:quantum_asymmetry}, respectively. The former, involves directly measuring the light scattered by mechanical motion and calibrating its intensity given a set of system parameters, while the latter requires comparing the emission and absorption rates of phonons from mechanical subsystem, and observing a non-classical asymmetry analogous to that seen in much earlier experiments with trapped atoms and ions~\cite{Diedrich1989,Jessen1992,Monroe1995}. Both methods are susceptible to laser phase noise, which is discussed in section~\ref{ss:laser_phase_noise}, through both noise induced heating~\cite{Rabl2009b}, and systematic errors in thermometry caused by noise \textit{squashing}~\cite{Poggio2007,Rocheleau2010}  and \textit{anti-squashing}. Finally, in Section~\ref{sec:experiments} we review recent laser cooling and sideband asymmetry measurements of optomechanical crystal cavities near the quantum ground state of their mechanical motion, and compare these results with the measured phase noise (see \ref{ss:broadband_measurement}) of the external cavity semiconductor diode lasers used in these experiements.

\section{Theory}\label{sec:theory}

\begin{figure}[ht]
\begin{center}
\includegraphics[width=\linewidth]{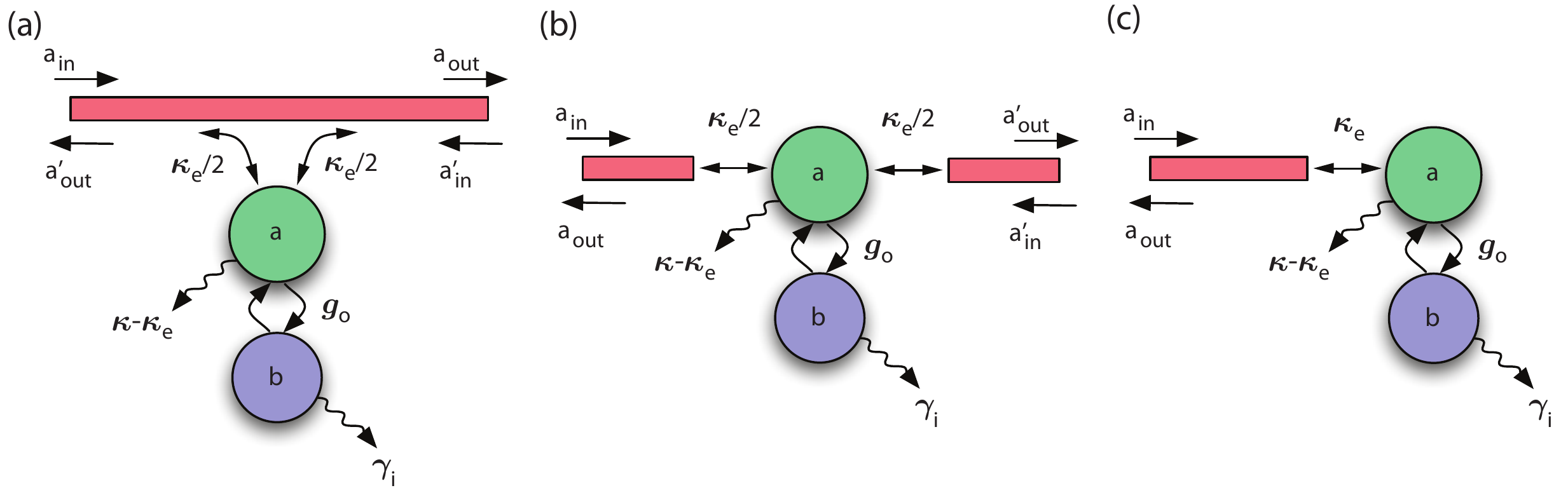}
\end{center}
\caption{\textbf{a}, Bi-directional evanescent coupling geometry, in which the ``transmitted'' field goes into the forward $a_\text{out}$ waveguide channel and the ``reflected'' field goes into the backward $a^{\prime}_\text{out}$ waveguide channel. This is the coupling geometry we will be focusing on in this work, in which the laser input channel is $a_\text{in}$ and the detection channel is the forward waveguide channel, $a_\text{out}$.  \textbf{b}, Double-sided end-fire coupling geometry.  This is the geometry one would have in a Fabry-Perot cavity.   Note that in this geometry what we call the ``transmission'' and ``reflection'' channels are typically opposite of that in the evanescent coupling geometry (a direct map between the two geometries would relate the ``reflected'' channel in (a) to the conventional transmission of a Fabry-Perot, for instance).  \textbf{c}, Single-sided end-fire couplng geometry.  This is the ideal measurement geometry, in which all of the optical signal that is coupled into the cavity can be, in principle, collected and detected in the $a_\text{out}$ channel.  In principle one does not have excess vacuum noise coupled into the optical cavity, only that from the laser input channel $a_\text{in}$.
 \label{fig:coupling_geometry}}
\end{figure}

The optomechanical interaction between a mechanical system and an optical field occurs through radiation pressure, a force proportional to the optical field intensity. This can be modeled by a Hamiltonian
\bea
H = \hbar \omega_o \opdagger{a}{} \op{a}{} + \hbar \omega_{m0} \opdagger{b}{} \op{b}{} + \hbar g \opdagger{a}{}\op{a}{}(\opdagger{b}{} + \op{b}{}),
\eea
with $\op{a}{}$ and $\op{b}{}$ the annihilation operators for photons and phonons in the system. In the presence of a laser emitting light at frequency $\omega_L$, it is convenient to work in an interaction frame where $\omega_o \rightarrow \Delta$ in the above Hamiltonian with $\Delta = \omega_o - \omega_L$. To incorporate the effect of the environment, we use the quantum-optical Langevin equations for the system~\cite{Gardiner1985,Collett1984,Gardiner2004},
\bea
&\dot{\op{b}{}}(t) = -\left(i\omega_{m0} + \frac{\gamma_i}{2}\right) \op{b}{} - ig \opdagger{a}{} \op{a}{} - \sqrt{\gamma_i}\op{b}{\mathrm{in}}(t)\nonumber
\eea
and
\bea
&\dot{\op{a}{}}(t) = -\left(i\Delta + \frac{\kappa}{2}\right) \op{a}{} - ig \op{a}{}(\opdagger{b}{}+\op{b}{}) -\sqrt{\kappa_e/2} \op{a}{\mathrm{in}}(t) - \sqrt{\kappa^\prime}\op{a}{\mathrm{in,i}}(t).
\eea
Here $\op{a}{\mathrm{in}}(t)$ and $\op{a}{\mathrm{in,i}}(t)$ are the quantum noise operators associated with extrinsic (input/output) and intrinsic (undetected) optical loss channels, respectively.  Here we assume bi-directional evanescent waveguide coupling (see Fig.~\ref{fig:coupling_geometry}) to the optical cavity in which the total extrinsic cavity loss rate into both directions of the coupling waveguide channel is $\kappa_e$, with the uni-directional input coupling rate being half that at $\kappa_e/2$.  The total optical cavity (energy) decay rate is given by $\kappa$, with $\kappa^\prime = \kappa - \kappa_e/2$ denoting all the optical loss channels which go undetected.  \footnote{In an idealized measurement, all photons lost by the optical cavity are lost to the detected channel so that $\kappa^\prime = 0$. In this ideal case, satisfied by the single-sided end-fire coupling geometry of Fig.~\ref{fig:coupling_geometry}c, the vacuum fluctuations from ports other than the detector port never enter the optical cavity. For a bi-directional coupling scheme we have $\kappa^\prime \geq \kappa_e/2$, and there is information lost in the backwards waveguide direction about the mechanical state.}  The noise operator $\op{b}{\text{in}}(t)$ arises from the coupling of the mechanical system to the surrounding bath degrees of freedom, which in most current systems resides in a high temperature thermal state with average bath occupancy $n_b \gg 1$.  

We linearize the equations about a large optical field intensity by displacing $\op{a}{} \rightarrow \alpha_0 + \op{a}{}$. Such an approximation is valid for systems such as ours where $g_0 \ll \kappa$, and the optical vacuum alone only marginally affects the dynamics of the system, i.e. the \textit{vacuum weak coupling} regime. All experimental systems to date are in this regime. For systems in the \textit{vacuum strong coupling} ($g_0 > \kappa$) regime, more elaborate treatments taking into account the quantum nature of the nonlinearity must be pursued~\cite{Ludwig2008,Rabl2011,Nunnenkamp2011}. In the Fourier domain the operators for the mechanical and optical modes are found to be 
\bea
&\op{b}{}(\omega) =  \frac{-\sqrt{\gamma_i}\op{b}{\mathrm{in}}(\omega)}{i(\omega_{m0}-\omega) + \gamma_i/2} -  \frac{i G(\op{a}{}(\omega) + \opdagger{a}{}(\omega))}{i(\omega_{m0}-\omega) + \gamma_i/2}\label{eqn:mech_fluct}
\eea
and
\bea
&\op{a}{}(\omega) =  \frac{-\sqrt{\kappa_e/2} \op{a}{\mathrm{in}}(\omega) - \sqrt{\kappa^\prime}\op{a}{\mathrm{in,i}} - iG(\op{b}{}(\omega) + \opdagger{b}{}(\omega))  }{i(\Delta-\omega) + \kappa/2}\label{eqn:opt_fluct},
\eea
respectively, where $G = g|\alpha_0|$.

Using equations (\ref{eqn:mech_fluct}-\ref{eqn:opt_fluct}) we arrive at the operator for the mechanical fluctuations,
\bea
\op{b}{}(\omega) =  \frac{-\sqrt{\gamma_i}\op{b}{\mathrm{in}}(\omega)}{i(\omega_m-\omega) + \gamma/2}~~~~~~~~~~~~~~~~~~~~~~~~~~~~~~~\nonumber\\
+ \frac{iG}{i(\Delta -\omega) + \kappa/2} \frac{\sqrt{\kappa_e/2} \op{a}{\mathrm{in}}(\omega) + \sqrt{\kappa^\prime} \op{a}{\mathrm{in,i}}(\omega)}{i(\omega_m-\omega) + \gamma/2}\nonumber\\
+ \frac{iG}{-i(\Delta + \omega) + \kappa/2} \frac{\sqrt{\kappa_e/2} \opdagger{a}{\mathrm{in}}(\omega) + \sqrt{\kappa^\prime} \opdagger{a}{\mathrm{in,i}}(\omega)}{i(\omega_m-\omega) + \gamma/2}, \label{eqn:b_inputs}
\eea
where $\omega_m=\omega_{m0}+\delta\omega_m$ is the optical spring shifted mechanical frequency and $\gamma = \gamma_i + \gamma_\mathrm{OM}$ is the optically damped (or amplified) mechanical loss-rate.  Expressions for the optical springing and damping terms are given by
\bea
&\delta\omega_m = |G|^2 \mathrm{Im}\left[ \frac{1}{i(\Delta-\omega_m)+\kappa/2} - \frac{1}{-i(\Delta+\omega_m)+\kappa/2} \right]\label{eq_spring}
\eea
and
\bea
&\gamma_{\mathrm{OM}} = 2|G|^2 \mathrm{Re}\left[ \frac{1}{i(\Delta-\omega_m)+\kappa/2} - \frac{1}{-i(\Delta+\omega_m)+\kappa/2} \right]\label{eq_damp},
\eea
respectively.  From these expressions it is evident that in the sideband-resolved regime the maximum optical damping occurs for a laser red-detuned from the optical cavity with $\Delta = \omega_m$, resulting in a damping rate of $\gamma_\mathrm{OM} \cong 4|G|^2/\kappa$. The ratio between this optical contribution to the mechanical damping and the intrinsic mechanical damping is the co-operativity, $C \equiv \gamma_\mathrm{OM}/\gamma_i$.

\subsection{Quantum-limited laser cooling and damping}\label{ss:quantum_limited_damping}

The expression for the noise power spectrum of the laser driven mechanical system can be calculated using eqn. (\ref{eqn:b_inputs}) for $\op{b}{}(\omega)$. More specifically we calculate $S_{bb}(\omega)$ (see \ref{app:mech_osc}), corresponding in the high-$Q$ regime to the ability of the mechanical system to \emph{emit} noise power into its environment~\cite{Clerk2010}.  The area under $S_{bb}(\omega)$ is the average mode occupancy of the mechanical quantum oscillator.  In the absence of the optical coupling to the mechanics ($G = 0$), the result in \ref{app:mech_osc} is obtained. Allowing for optical coupling and including the optical noise terms, we arrive at an expression involving the correlations $\avg{\opdagger{a}{\mathrm{in}}(\omega)\op{a}{\mathrm{in}}(\omega^\prime)}$, $\avg{\opdagger{a}{\mathrm{in,i}}(\omega)\op{a}{\mathrm{in,i}}(\omega^\prime)}$, $\avg{\op{a}{\mathrm{in}}(\omega)\opdagger{a}{\mathrm{in}}(\omega^\prime)}$ and $\avg{\op{a}{\mathrm{in,i}}(\omega)\opdagger{a}{\mathrm{in,i}}(\omega^\prime)}$ that must be calculated from the properties of the optical bath. Assuming that our source of light is a pure coherent tone, and thus the optical bath is in a vacuum state, as is approximately the case in many optical experiments, the former two correlations can be set to zero, while the latter two give $\delta(\omega+\omega^\prime)$.  As described in \ref{app:mech_osc}, for the mechanical system which is in contact with a thermal bath of occupancy $n_b$, we have noise input correlations of $\avg{\op{b}{\mathrm{in}}(\omega)\opdagger{b}{\mathrm{in}}(\omega^\prime)} = (n_b+1)\delta(\omega+\omega^\prime)$ and $\avg{\opdagger{b}{\mathrm{in}}(\omega)\op{b}{\mathrm{in}}(\omega^\prime)} =n_b \delta(\omega+\omega^\prime)$.  The expression for $S_{bb}(\omega)$ is then found to be,
\bea
S_{bb}(\omega) = \frac{\gamma {n_f}(\omega)}{(\omega_m + \omega)^2 + (\gamma/2)^2},\label{eqn:Sbb}
\eea
where $n_f(\omega)$, the back-action modified phonon occupation number, is given by
\bea
n_f(\omega) = \frac{\gamma_i n_b}{\gamma} + \frac{|G|^2\kappa}{\gamma}\frac{1}{(\Delta-\omega)^2 + (\kappa/2)^2}.
\eea
In the \textit{driven weak-coupling} regime $(\kappa \gg \gamma)$, the mechanical lineshape is not strongly modified from that of a Lorentzian, and $n_f(\omega)$ can simply be replaced by $n_f(-\omega_m)$ in eqn. (\ref{eqn:Sbb}).
An input laser beam tuned a mechanical frequency red of the cavity for optimal laser cooling ($\Delta = \omega_m$),  results in a back-action modified average mechanical mode occupation number equal to
\bea
\nbar\arrowvert_{\Delta=\omega_m} = \frac{\gamma_i n_b}{\gamma} + \frac{\gamma_\mathrm{OM}}{\gamma}\left(\frac{\kappa}{4\omega_m}\right)^2\label{eqn:quantum_limit_damping}.
\eea
The term $n_{qbl}\equiv (\kappa/4\omega_m)^2$ is the quantum limit on back-action cooling, as derived in \cite{Wilson-Rae2007,Wilson-Rae2008} using master-equation methods and in \cite{Marquardt2007} by taking into account the spectral density of the optical back-action force.
This small (in the good cavity limit) residual heating comes from the non-resonant scattering of red pump photons, to one mechanical frequency lower, or a total of $2\omega_m$ detuned from the optical cavity.

We note briefly that for other laser detunings different back-action occupancies are achieved, such as
\bea
&\nbar\arrowvert_{\Delta = 0} = n_b + \frac{4 |G|^2}{\gamma_i\kappa}\left(\frac{\kappa}{2\omega_m}\right)^2 \label{eqn:quantum_limit_damping_0}
\eea
and
\bea
&\nbar\arrowvert_{\Delta = -\omega_m} = \frac{\gamma_i n_b}{\gamma} + \frac{|\gamma_\mathrm{OM}|}{\gamma},\label{eqn:quantum_limit_damping_-}
\eea
where again the resolved-sideband limit is assumed, and for $\Delta=-\omega_m$ one has amplification of the mechanical motion with $\gamma_\mathrm{OM} \cong -4|G|^2/\kappa$.

\subsection{Thermometry with the cooling beam}\label{ss:simple_therm}

One of the simplest methods of inferring the mechanical mode occupancy is to detect the imprinted mechanical motion on the cooling laser beam itself.  Upon transmission through the cavity-optomechanical system, the laser cooling beam, typically detuned to  $\Delta = \omega_m$ in the resolved sideband regime, preferentially picks up a blue-shifted sideband at frequency $\omega_L + \omega_m$ ($\approx \omega_o$) due to removal of phonon quanta from the mechanical resonator (anti-Stokes scattering).  Upon detection with a photodetector, the beating of the anti-Stokes sideband with the intense cooling tone produces an electrical signal at the mechanical frequency~\cite{Schliesser2006,Park2009,Chan2011}. By careful calibration and accurate measurement of the magnitude of this signal, and through independent measurements of other system parameters such as $g$, $|\alpha_0|$, $\kappa_e$, and $\kappa$, the mechanical resonator's average phonon number occupancy can be inferred. 

The optical fluctuations in the transmitted laser cooling beam at the output port of the optomechanical cavity are given approximately in the sideband resolved regime by
\bea
\op{a}{\mathrm{out}}(\omega)\arrowvert_{\Delta=\omega_m} \approx t(\omega;\Delta)\op{a}{\mathrm{in}}(\omega) + n_{\mathrm{opt}}(\omega;\Delta)\op{a}{\mathrm{in,i}}(\omega) + s_{12}(\omega;\Delta)\op{b}{\mathrm{in}}(\omega), \label{eqn:aout_redside}
\eea
where $t$, $n_\mathrm{opt}$, and $s_{12}$ are the scattering matrix elements evaluated for a laser cooling beam of red-detuning $\Delta=\omega_m$ (see \ref{app:scattering_matrix}). This expression is derived using eqns. (\ref{eqn:mech_fluct}-\ref{eqn:opt_fluct}) and input-output boundary condition $\op{a}{\mathrm{out}}(\omega) = \op{a}{\mathrm{in}}(\omega) + \sqrt{\kappa_e/2}\op{a}{}(\omega)$. Expressions of this form have been used previously to analyze the propagation of light and sound through an optomechanical cavity in the context of state transfer~\cite{Safavi-Naeini2011a, Hill2012}.  In this case we have simplified eqn.\ (\ref{eqn:aout_redside}) by ignoring input noise terms from the creation operators $\opdagger{a}{\mathrm{in}}(\omega)$ and $\opdagger{a}{\mathrm{in,i}}(\omega)$. These terms gives rise to the quantum-limit of laser cooling found in eqn. (\ref{eqn:quantum_limit_damping}) above, but insignificantly modify the optically transduced signal of the mechanical motion as long as $\nbar \gg n_{qbl}$.  

The strong cooling laser tone beats with the optical noise sidebands, generating a photocurrent proportional to $\op{a}{\mathrm{out}}(t) + \opdagger{a}{\mathrm{out}}(t)$,  
\bea
\op{I}{}(\omega)\arrowvert_{\Delta=\omega_m} &=&  t(\omega)\op{a}{\mathrm{in}}(\omega) + n_{\mathrm{opt}}(\omega)\op{a}{\mathrm{in,i}}(\omega) + s_{12}(\omega)\op{b}{\mathrm{in}}(\omega) +\nonumber\\
&&+\mathrm{h.c.}(-\omega)
\nonumber
\eea
where $\mathrm{h.c.}(-\omega)$ is a convenient short-hand ($f(\omega) + (f(-\omega))^\dagger = f(\omega) +  \mathrm{h.c.}(-\omega)$). The resulting photocurrent power spectral density as read out from a spectrum analyzer is given by
\bea
S_{II}(\omega)\arrowvert_{\Delta=\omega_m} &=& \int_{-\infty}^{\infty} \mathrm{d}\omega^\prime~ \avg{\opdagger{I}{}(\omega)\op{I}{}(\omega^\prime)} \nonumber\\
 &=& \int_{-\infty}^{\infty} ~ t(\omega)t(-\omega^\prime)^\ast \avg{\op{a}{\mathrm{in}}(\omega)\opdagger{a}{\mathrm{in}}(\omega^\prime)}\nonumber\\
&&+ n_\mathrm{opt}(\omega)n_\mathrm{opt}(-\omega^\prime)^\ast \avg{\op{a}{\mathrm{in,i}}(\omega)\opdagger{a}{\mathrm{in,i}}(\omega^\prime)}\nonumber\\
&&+ s_{12}(\omega)s_{21}(-\omega^\prime)^\ast \avg{\op{b}{\mathrm{in}}(\omega)\opdagger{b}{\mathrm{in}}(\omega^\prime)}\nonumber\\
&&+ s_{12}(-\omega)s_{21}(\omega^\prime)^\ast \avg{\opdagger{b}{\mathrm{in}}(\omega)\op{b}{\mathrm{in}}(\omega^\prime)}\mathrm{d}\omega^\prime,
\eea
where we have assumed the same optical (vacuum) and mechanical (thermal) noise correlations as above in evaluating eqn. (\ref{eqn:Sbb}).  Using the normalization property of the scattering matrix coefficients ($|t(\omega)|^2 + |n_\mathrm{opt}(\omega)|^2 + |s_{12}(\omega)|^2 = 1$), we find the simplified expression
\bea
S_{II}(\omega)\arrowvert_{\Delta=\omega_m} &=& 1 + n_b(|s_{12}(\omega)|^2 + |s_{12}(-\omega)|^2).
\eea
Substituting for the expression of the phonon-photon scattering element $s_{12}(\omega)$ given in \ref{app:scattering_matrix} for $\Delta=\omega_m$ yields 
\bea
S_{II}(\omega)\arrowvert_{\Delta=\omega_m} &=&  1 + \frac{\kappa_e}{2\kappa} \frac{8|G|^2}{\kappa} \bar{S}_{bb}(\omega;\nbar),\label{eqn:SII}
\eea
for the transduced noise power spectral density, where $\nbar$ is the actual mechanical mode occupancy including back-action effects of the cooling laser (see eqn.~(\ref{eqn:quantum_limit_damping})).

Several points are worth mentioning regarding this expression. Firstly, the signal to noise goes as the coupling efficiency $\eta = \kappa_e/2\kappa$. Secondly, the detected signal is proportional to $\avg{\opdagger{b}{}\op{b}{}}$ as opposed to $\avg{\opdagger{b}{}\op{b}{}}+1/2$, and so the resulting signal is exactly what would be expected classically, vanishing as the temperature and phonon occupation go to zero. In other words, this measurement is insensitive to the zero-point motion of the resonator. The spectral density $S_{bb}(\omega,\nbar)$, represents the ability of the mechanical system to \textit{emit} energy~\cite{Clerk2010}. By tuning the laser to $\Delta=\omega_m$ in the sideband-resolved regime it is exceedingly unlikely for the tone to drive the mechanics (through Stokes scattering), and so we gain little information about how the mechanical system \textit{absorbs} energy from the optical bath. Finally, we note that equation (\ref{eqn:SII}) is general and holds for both low and high cooperativity.

\subsubsection{Intepretation as quantum noise squashing}\label{ss:interp_noisesquashing}

Though the scattering matrix formulation provides a consistent and systematic way of deriving the form of the detected signals, it does so by elimination of the position operator from the equations. It is interesting to reinterpret the experiment as a measurement of the position of the mechanical system~\cite{Clerk2010}, and we attempt to do so here\footnote{A much more thorough treatment of the implications of quantum back-action and measurement theory in this type of system is presented in a recent work by Khalili, et al.~\cite{Khalili2012}}. For simplicity, the perfect coupling condition is assumed, i.e. $\kappa_e/2 = \kappa$. The output signal is then given by
\bea
\op{a}{\mathrm{out}}(\omega)\arrowvert_{\Delta=\omega_m}  \approx -\op{a}{\mathrm{in}}(\omega)  - \frac{i2G}{\sqrt{\kappa}} \op{b}{}(\omega) \nonumber
\eea
The normalized heterodyne current is found to be
\be
\op{I}{}(t)\arrowvert_{\Delta=\omega_m} = - i \op{a}{\mathrm{in}}(t) + i \opdagger{a}{\mathrm{in}}(t) + \frac{2G}{\sqrt{\kappa}} (\op{b}{}(t) +  \opdagger{b}{}(t))
\ee
and so it would seem that the signal $\op{I}{}(t)$ is composed of optical shot-noise and a component which is proportional to $\op{x}{}$, making $S_{II} = 1 +\text{const}\times  \avg{\op{x}{}^2}$. This however contradicts the above derivations which show that $S_{II} = 1 +\text{const}\times \nbar$ for $\Delta=\omega_m$. The inconsistency comes after careful calculation of the correlation function $\avg{\op{I}{}(t+\tau)\op{I}{}(t)}$. In fact, $\opdagger{a}{\mathrm{in}}(t)$ and $\op{b}{}(t)$ are correlated, and the view that the shot-noise simply creates a constant noise floor is incorrect. Proper accounting for the correlations (see \ref{app:quantum_noise_squashing}) leads us again to eqn.~(\ref{eqn:SII}), showing that the measured quantity is $\bar{S}_{bb}$, and the area of the detected spectrum is proportional to $\nbar$. 

The blue-side driving with $\Delta=-\omega_m$ causes the opposite effect, i.e. quantum noise anti-squashing. The squashing and anti-squashing are signatures of quantum back-action. This effect is in spirit similar to classical noise squashing which we study in Section~\ref{ss:squashing}, where correlations between the noise-induced motion and \textit{classical} noise of the detection beam destructively interfere at the photodetector. It is important to note that this signature of quantum back-action does not involve detection of quantum back-action heating, and can be apparent at arbitrarly low powers, far below that required to reach the standard quantum limit.

\subsection{Motional sideband asymmetry thermometry}\label{ss:quantum_asymmetry}

An alternate method of measuring the temperature of the mechanical subsystem, one which uses the mechanical zero-point motion to self-calibrate the measured phonon occupancy, involves comparing the measured signal from a weak probe beam (low cooperativity) at both $\Delta=\pm \omega_m$ in the sideband resolved regime~\cite{Safavi-Naeini2012}. In such experiments, the mechanics can be either laser cooled with a different laser and/or optical cavity mode, or the system can be cryogenically pre-cooled to a temperature which requires no further cooling to approach the quantum ground state. As the optical read-out beam can be arbitrarily weak in such measurements, it only marginally affects the dynamics of the mechanical  system~\cite{Schliesser2009,Groeblacher2009a,Safavi-Naeini2012}. By working at low read-out beam power, such that the optically-induced damping and amplification rates are much smaller than the bare mechanical linewidth, optical back-action by the probe beam only minimally affects the dynamics of the mechanical system and measurements can be taken at detunings both red ($\Delta=\omega_m$) and blue ($\Delta=-\omega_m$) of the cavity without triggering any optomechanical instabilities~\cite{Marquardt2006}. Operating in the resolved sideband regime allows for the separate cavity filtering of the Stokes and anti-Stokes motionally induced sidebands on the probe beam, which are respectively proportional to $\nbar + 1$ and $\nbar$. It can be shown that the additional vacuum contribution to the Stokes scattering, which provides the intrinsic calibration for $\nbar$, arises in these measurements equally from the shot noise on the probe laser and zero-point motion of the mechanical resonator~\cite{Khalili2012}.  We will also see in the following sections, that such a measurement at both $\Delta=\pm\omega_m$ can provide additional resilience to systematic errors from non-idealities such as laser phase noise.

We derive here the blue-detuned ($\Delta = -\omega_m$) result analogous to the red-detuned ($\Delta = \omega_m$) laser cooling case given above in eqn.(\ref{eqn:SII})). In the sideband-resolved regime, the approximations that led to eqn. (\ref{eqn:aout_redside}), lead to a similar expression in the case of $\Delta=-\omega_m$ for the electromagnetic field output from the optomechanical cavity 
\bea
\op{a}{\mathrm{out}}(\omega)\arrowvert_{\Delta=-\omega_m} \approx t(\omega;\Delta)\op{a}{\mathrm{in}}(\omega) + n_{\mathrm{opt}}(\omega;\Delta)\op{a}{\mathrm{in,i}}(\omega) + s_{12}(\omega;\Delta)\opdagger{b}{\mathrm{in}}(\omega). \label{eqn:aout_blue}
\eea
where we have neglected the terms proportional to the photon noise creation operators as their effect is again minimal on the optically transduced signal of the mechanical motion. Such a scattering relation, whose exact form is shown in the \ref{app:scattering_matrix}, allows the optomechanical system to act as an amplifier, and has been studied experimentally at microwave~\cite{Massel2011a} and optical frequencies~\cite{Safavi-Naeini2011}, and studied more generally in the context of optomechanics by Botter, et al.~\cite{Botter2012}. The scattering elements satisfy the equation $|t(\omega)|^2 + |n_\mathrm{opt}(\omega)|^2 + |s_{12}(\omega)|^2 = 1$, which along with the standard bath correlation relations used above, allows us to write
\bea
S_{II}(\omega)\arrowvert_{\Delta=-\omega_m} &=& 1 + (n_b+1)(|s_{12}(\omega;\Delta)|^2 + |s_{12}(-\omega;\Delta)|^2) \\
&=&  1 + \frac{\kappa_e}{2\kappa} \frac{8|G|^2}{\kappa} \bar{S}_{b^\dagger b^\dagger}(\omega;\nbar),\label{eqn:SII_blueside}
\eea
where $\nbar$ is the actual mode occupancy including back-action of the laser input (see eqn.~(\ref{eqn:quantum_limit_damping_-})).  As before, the signal lies on top of a flat shot noise background of unity, and is proportional to the detection efficiency $\eta$ and the measurement rate $\gamma_\text{OM}$.  Now, however, the signal is proportional to the creation operator spectral density $\bar{S}_{b^\dagger b^\dagger}(\omega;\nbar)$, which itself is proportional to $\nbar + 1$. The spectral density $\bar{S}_{b^\dagger b^\dagger}(\omega;\nbar)$ can be interpreted as the mechanical system's ability to \textit{absorb} energy~\cite{Clerk2008}, which even at zero temperature (occupation) can absorb energy through spontaneous scattering process which arises due to the zero-point motion of the mechanical resonator.

For a constant laser driving power the optomechanical damping and amplification rates for detuning $\Delta=\pm\omega_m$ are equal in magnitude but opposite in sign, with $\gamma_\pm \equiv \gamma_i \pm |\gamma_\text{OM}|$, where $|\gamma_\text{OM}| \cong 4|G|^2/\kappa$ in the sideband resolved, weak coupling regime. Weak probing entails using a probe intensity such that $|\gamma_\text{OM}| \ll \gamma_i$, or $C_r \ll 1$, where we define $C_r\equiv |\gamma_\text{OM}|/\gamma_i$ as the read-out beam cooperativity. In this limit the mechanical mode occupation numbers for $\Delta=\pm \omega_m$ detunings are given approximately by $\nbar_\pm \cong \gamma_i n_b / \gamma_\pm$, where $n_b$ is the mechanical mode occupancy in absence of the probe field\footnote{Referring to eqns.~(\ref{eqn:quantum_limit_damping}) and (\ref{eqn:quantum_limit_damping_-}), this is an accurate relation if $C_r \ll n_b$.}. Denoting the integrated area under the Lorentzians in eqns. (\ref{eqn:SII}) and (\ref{eqn:SII_blueside}) as $I_+$ and $I_-$, respectively, we find a relation between their ratios and the read-out cooperativity which provides a quantum calibration of the unperturbed thermal occupancy~\cite{Safavi-Naeini2012}:
\bea
\eta \equiv \frac{I_-/I_+}{1+C_r}  - \frac{1}{1-C_r} = \frac{1}{n_b}.\label{eqn:eta}
\eea


\subsection{Laser phase noise}\label{ss:laser_phase_noise}

Although various other noise sources (laser intensity noise, internal cavity noise~\cite{Rocheleau2010}, etc.) can be treated similarly, here we focus on laser phase noise as it is typically the most important source of nonideality in the laser cooling and thermometry of cavity opto-mechanical systems. The effect of phase noise on optomechanical systems has already been studied at great depth in the context of heating~\cite{Rabl2009b} of the mechanical resonator and entanglement~\cite{Phelps2011,Abdi2011,Ghobadi2011} of light and mechanics.  However, laser light often acts as both a means by which the mechanical system is cooled as well as its temperature measured, and thus laser noise can effect both the true and inferred mechanical mode occupancy.  Here we complement previous studies of laser noise heating with a unified analysis that also quantifies the effects of quantum and classical (phase) laser noise on the optically-transduced mechanical mode spectra.  

The optical laser field amplitude input to the optomechanical system, in a rotating reference frame at frequency $\omega_L$ and in units of $\sqrt{\text{photons/s}}$, we denote by $E_0$. Due to processes internal to the laser, some fundamental in nature, others technical, this amplitude undergoes random phase fluctuations which are captured by adding a random rotating phase factor~\cite{Meschede2004}
\bea
E_0(t) = |E_0| e^{i\phi(t)}.
\eea
As long as the phase fluctuations are small, we expand this expression to first order yielding $E_0(t)\approx |E_0|(1+i\phi(t)+O(\phi^2))$~\cite{Rabl2009b}. Then
\bea
\avg{E_0^\ast(\tau)E_0(0)} &=& |E_0|^2 \left(1+\avg{\phi(\tau)\phi(0)} \right)\nonumber.
\eea
In this way, we can express the noise power spectral density of the optical field amplitude, $S_{EE}(\omega)$, as
\bea
S_{EE}(\omega)&=& |E_0|^2(2\pi\delta(\omega) + \bar{S}_{\phi\phi}(\omega)) \label{eqn:S_EE_Sphiphi},
\eea
where we've also used the realness of $\phi(t)$ to set $S_{\phi\phi}(\omega) = \bar{S}_{\phi\phi}(\omega)$.

%
This relates the phase noise power spectral density to the optical power spectum of the noisy laser beam, with the optical power away from the carrier at $\omega=0$ due to phase noise.  This phase noise can then be taken into account as an additional noise input to the cavity, 
\bea
\op{a}{\mathrm{in,tot}}(\omega) = \op{a}{\mathrm{in}}(\omega) + {a}_{\mathrm{in},\phi}(\omega),
\eea
where $a_\mathrm{in,\phi}(\omega)$ is a stochastic input with $\avg{a^{\dagger}_\mathrm{in,\phi}(\omega)a_\mathrm{in,\phi}(\omega^\prime)} = S_{EE}(\omega)\delta(\omega+\omega^\prime)$ (here the averages used for correlation functions correspond to classical ensemble averages and $a^{\dagger}_\mathrm{in,\phi}(\omega)\equiv (a_\mathrm{in,\phi}(-\omega))^{*}$). 

There is, however, an additional subtlety when performing mechanical mode thermometry with a laser beam affected by phase noise; correlations between the positive and negative frequency components of the phase noise can cause cancellations in the optically transduced signal, and therefore must be carefully taken into account.  For example, for a pure sinusoidal tone phase modulated onto a laser we have,
\bea
E_0 e^{i \phi (t)} &\simeq& E_0 \left( e^{i \beta_{\mathrm{c}} \cos \omega t+i \beta_{\mathrm{s}} \sin \omega t} \right)\nonumber \\
& \simeq& E_0 \left(1 + \frac{1}{2} (\beta_{\mathrm{s}} + i \beta_{\mathrm{c}} ) e^{i \omega t} - \frac{1}{2} (\beta_{\mathrm{s}} - i\beta_{\mathrm{c}} ) e^{-i \omega t} \right). \nonumber
\eea
The positive and negative frequency optical sideband amplitudes are negative complex conjugates of one another. More generally, the positive and negative frequency components of the noisy optical input have the following relation for an optical signal with phase noise,
\bea
a^{(-)}_\mathrm{in,\phi}(\omega) = -\left(a^{(+)}_\mathrm{in,\phi}(-\omega) \right)^{\ast}
\eea
where $a^{(-)}_\mathrm{in,\phi}(\omega) = \theta(-\omega) a_\mathrm{in,\phi}(\omega)$, and $a^{(+)}_\mathrm{in,\phi}(\omega) = \theta(\omega) a_\mathrm{in,\phi}(\omega)$. The total phase noise signal can then expressed as $a_\mathrm{in,\phi}(\omega) = a^{(+)}_\mathrm{in,\phi}(\omega)  - a^{(+)\dagger}_\mathrm{in,\phi}(-\omega)$.  For calculations that follow, this explicit separation of positive and negative frequency phase noise components is useful in simplifying calculations of the optically transduced mechanical motion.  In terms of positive frequency phase noise components only then, we have
\bea
\avg{a^{(+)\dagger}_\mathrm{in,\phi}(\omega)a^{(+)}_\mathrm{in,\phi}(\omega^\prime)} = S_{EE}(\omega)\delta(\omega+\omega^\prime)\theta(\omega) \label{eqn:corr_a+},
\eea
with a similar relation holding for the negative frequency components of the phase noise input and the negative frequency optical power spectrum. 

\subsubsection{Heating}

To find the actual thermal occupation of the mechanical system in the presence of phase noise on the laser cooling beam we use once again eqn.~(\ref{eqn:b_inputs}) for $\op{b}{}(\omega)$, replacing $\op{a}{\mathrm{in}}(\omega)$ with $\op{a}{\mathrm{in,tot}}(\omega)$ which includes the classical phase noise on the input laser. From the non-zero correlation $\avg{\opdagger{a}{\mathrm{in,tot}}(\omega)\op{a}{\mathrm{in,tot}}(\omega^\prime)} = S_{EE}(\omega) \delta(\omega+\omega^\prime)$ for $\omega,\omega^\prime >  0$, we find another source of noise phonons, in addition to those coming from the thermal bath and quantum back-action of the laser light. This is expressed as new terms proportional to $S_{EE}(\omega)$ in the noise spectrum of the mechanical motion given by eqn.~(\ref{eqn:Sbb}),
\bea
&S_{bb}(\omega) = \frac{\gamma {n_{f,\phi}}(\omega)}{(\omega_m + \omega)^2 + (\gamma/2)^2},
\eea
where
\bea
n_{f,\phi}(\omega)\arrowvert_{\Delta=\omega_m} &=& \frac{\gamma_i n_b}{\gamma} + \frac{|G|^2\kappa}{\gamma}\frac{1+(\kappa_e/2\kappa)S_{EE}(\omega)}{(\Delta-\omega)^2 + (\kappa/2)^2}\nonumber\\
&& + \frac{|G|^2\kappa}{\gamma}\frac{(\kappa_e/2\kappa)S_{EE}(\omega)}{(\Delta+\omega)^2 + (\kappa/2)^2}.
\eea
As before, assuming a high mechanical $Q$-factor and the driven weak-coupling regime $(\kappa \gg \gamma)$, we substitute $n_{f,\phi}(-\omega_m)$ for $n_{f,\phi}(\omega)$ and relate it to the average mode occupancy in the presence of laser phase noise, $\nbar_{\phi}$,
\bea
\nbar_{\phi}\arrowvert_{\Delta=\omega_m} = \frac{\gamma_i n_b}{\gamma} + \frac{\gamma_\mathrm{OM}}{\gamma}\left[\left(\frac{\kappa}{4\omega_m}\right)^2+\left(\frac{\kappa_e}{2\kappa}\right)n_\phi\right]\label{eqn:quantum_phasenoise_limit_damping},
\eea
where we have defined $n_\phi \equiv S_{EE}(\omega_m)$~\cite{Rabl2009b}.

The additional phase noise heating in equation~(\ref{eqn:quantum_phasenoise_limit_damping}) can be understood as the product of the number of noise photons present in the light field at a mechanical frequency detuned from the central laser frequency ($n_\phi$), multiplied by the efficiency with which they are coupled into the cavity ($\kappa_e/2\kappa$), and finally multiplied by the efficiency with which the light field couples to the mechanics ($\gamma_\text{OM}/\gamma$).  For detuning $\Delta=\omega_m$ used in resolved sideband laser cooling, the optomechanical system only samples the input laser phase noise at a mechanical frequency blue of the central laser frequency, and thus the relationship between the negative and positive frequency components of the laser phase noise has no role to play in heating in the sideband-resolved regime (as we will see below, this is not the case in the thermometry). 

\subsubsection{Effect on calibrated cooling beam thermometry}\label{ss:squashing}

\begin{figure}[ht]
\begin{center}
\includegraphics[width=0.65\linewidth]{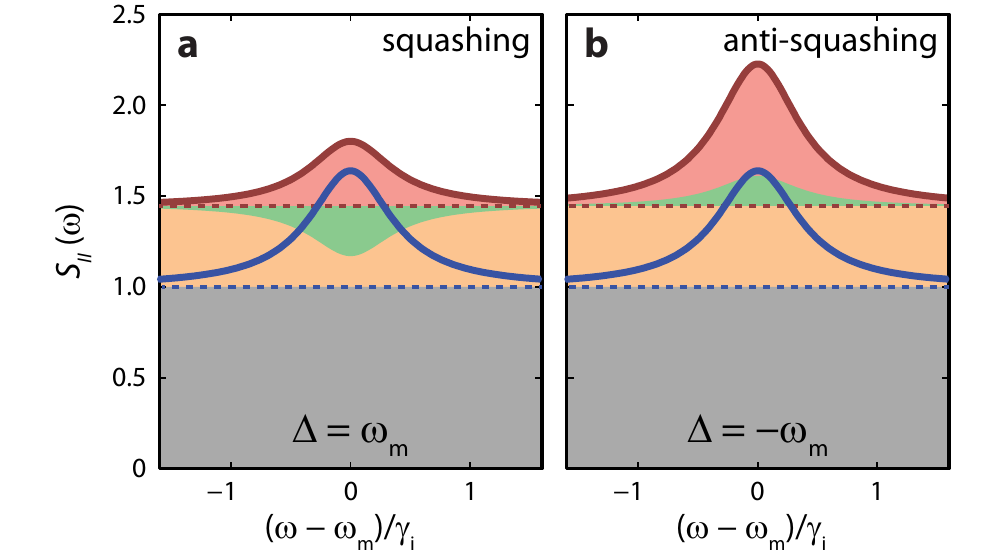}
\end{center}
\caption{\textbf{a}, and \textbf{b} show the spectral density of the detected signal, $S_{II}(\omega)$ for a large phonon occupation $\bar{n} \gg 1$, about the mechanical frequency. The contributions from phase noise (in orange), thermal brownian motion (in blue) and shot-noise (in grey), are highlighted. In \textbf{a}, the effect of noise squashing, arising from the cancellation of phase noise near the mechanical frequency is apparent (green shaded area). Without phase noise, the spectrum in blue would be detected, with an area corresponding to the actual phonon population. The dip from the phase noise background reduces the area under this curve, causing the area to be underestimated. In \textbf{b} the spectra from blue-side detection are presented. Here we note that the phase noise is amplified, causing a larger temperature. Though \textbf{a} and \textbf{b} are for large phonon occupation numbers, an asymmetry still appears due to classical laser phase noise.
 \label{fig:squashing_anti_squashing}}
\end{figure}

Since direct photodetection is sensitive to the total field intensity and not the phase, laser light with phase noise detected without first passing through a system with frequency dependent transmission will fail to exhibit any fluctuations in excess of shot-noise. Ignoring the mechanical part of the system for the moment, a laser detuned from an optical cavity by $\Delta$ will, however, cause the phase noise of the laser near frequency $\Delta$ to appear as an increased noise floor level in the photodetected spectrum of the transmitted optical signal.  The noise from the mechanical system undergoing random motion appears as a Loretzian peak on top of this noise floor in the photocurrent spectrum. 
Depending on the exact experimental geometry, i.e. whether the measurement is done on reflection or transmission and whether the probe laser is red- or blue-detuned from the cavity, the amplitude of this lorentzian signal above the raised noise floor, can be in excess or below that expected in an ideal measurement lacking laser phase noise. At very high relative noise levels, it is even possible for the signal peak to invert, and become a dip in the noise floor.
 Such an effect has been called noise squashing~\cite{Poggio2007}, and results from correlations in the input laser noise and the optically transduced mechanical motion noise due to radiation pressure fluctuations stemming from the same input laser noise.  

Figures~\ref{fig:squashing_anti_squashing}a and b display a model of the photocurrrent noise power spectrum for the transmitted laser field of an evanescently-coupled cavity-optomechanical geometry (Fig.~\ref{fig:coupling_geometry}a) for red ($\Delta=\omega_m$) and blue ($\Delta=-\omega_m$) laser detuning from the optical cavity resonance, respectively.  The shaded orange area denotes the part of the photocurrent noise power spectrum that is generated due to phase noise on the probe laser. It can be seen that the aforementioned interference effect, in the transmission geometry considered here, causes a dip (shaded green) in the phase noise background for red-side laser driving and a peak for blue-side laser driving\footnote{In the case of detecting in the reflection channel for this geometry, the reverse is seen, with red-side laser detuning resulting in anti-squashing and blue-side detuning resulting in squashing.}. Note that, over a broader bandwidth than that shown in Fig.~\ref{fig:squashing_anti_squashing}, the phase noise contribution to the photocurrent noise spectrum is modulated by the optical cavity lineshape, $\kappa \gg \gamma$.

Formally, the photocurrent noise power spectrum of Fig.~\ref{fig:squashing_anti_squashing} can be derived by considering the properties of a \emph{driven} cavity-optomechancal system.  The transmission and reflection of light by a laser driven optomechanical cavity has been of interest for a variety of switching and buffering applications~\cite{Chang2011}, and displays physics analogous to electromagnetically induced transparency (EIT)~\cite{Agarwal2010,Weis2010,Teufel2011a,Safavi-Naeini2011} and electromagnetically induced absorption and amplification (EIA)~\cite{Massel2011a,Botter2012,Hocke2012}.  Considering the evanescently coupled geometry of Fig.~\ref{fig:coupling_geometry}a, the reflection coefficient of (weak) probe light at frequencies $\omega$ from an intense laser drive tone at frequency $\omega_L = \omega_o \mp \Delta$, is given by
\bea \label{eq:r_omega_red}
r^\pm(\omega) &=& -\frac{\kappa_e/2}{i(\Delta \mp \omega) + \kappa/2 + \frac{|G|^2}{i(\omega_m-\omega)\pm\gamma_i/2}}. \label{eqn:refl_eit_eia}
\eea
Here reflection is into the backwards waveguide direction ($a_\text{out,-}$ of Fig.~\ref{fig:coupling_geometry}a).  The transmission coefficient into the output channel in the forward waveguide direction is $t^\pm(\omega) = 1 + r^\pm(\omega)$.  

The photocurrent of the detected signal in the forward waveguide direction output due to laser phase noise present at the input is,
\bea
I_\phi(\omega) &=& t(\omega)a_\mathrm{in,\phi}(\omega)  + \mathrm{h.c.}(-\omega),
\eea
which, taking into account the above relation between the transmission and reflection coefficients and the correlation between the positive and negative frequency components of the phase noise, yields in terms of the positive phase noise components only
\bea
I_\phi(\omega)&=&r(\omega)a^{(+)}_\mathrm{in,\phi}(\omega) +  \mathrm{h.c.}(-\omega).
\eea
Using eqns. (\ref{eqn:SAA_omega_integral}) and (\ref{eqn:corr_a+}), the resulting photocurrent noise power spectrum in the output channel due to laser phase noise is then calculated to be:
\bea
S_{I_\phi I_\phi} (\omega)  = \left( |r(\omega)|^2\theta(\omega) + |r(-\omega)|^2\theta(-\omega) \right) S_{EE}(\omega).
\eea
Evaluating the reflection coefficient at laser detuning $\Delta=\pm\omega_m$ yields for the full expression for a sideband-resolved system in the driven weak-coupling regime,
\bea
\frac{S^\pm_{I_\phi I_\phi} (\omega)}{S_{EE}(\omega)} &=&  \left(\frac{\kappa_e}{\kappa} \right)^2 \nonumber\\&&\mp \left(\frac{\kappa_e}{\kappa} \right)^2 \frac{|\gamma_\text{OM}|}{2} \frac{\gamma_i \pm |\gamma_\text{OM}/2|}{(\omega_m \mp \omega)^2 + (\gamma/2)^2}.\label{eq:I_phi_noise_power}
\eea

The above expression in eqn.~(\ref{eq:I_phi_noise_power}) is that plotted in Fig.~\ref{fig:squashing_anti_squashing}, showing the laser phase noise contribution to the measured output noise power power spectrum. Adding this noise spectrum with detuning set to $\Delta=\omega_m$, to that found in eqn.~(\ref{eqn:SII}) for the laser cooling beam output noise spectrum in the absence of classical laser noise, we find for the total transduced noise power spectral density near the mechanical resonance in the large cooperativity limit ($\gamma_i \ll \gamma_\mathrm{OM}$),
\bea
S_{II}(\omega)\arrowvert_{\Delta=\omega_m} &=&  1 +  \left(\frac{\kappa_e}{\kappa}\right)^2 n_\phi + \frac{\kappa_e}{2\kappa} \frac{8|G|^2}{\kappa} \bar{S}_{bb}(\omega; n_\mathrm{inf}),\label{eqn:SII_phasenoise}
\eea
where $n_\mathrm{inf} = \gamma_i n_b / \gamma - \kappa_e n_\phi/2\kappa$ and $n_b$ is mechanical mode occupancy in the absence of the cooling beam.  This is the noise output power spectral density in experiments in which the laser cooling beam is also used for transduction/thermometry of the mechanical mode.  Laser phase noise on the cooling beam, then, not only adds additional heating of the mechanical resonator (as captured by eqn.~(\ref{eqn:quantum_phasenoise_limit_damping})), but the naively inferred phonon number represented by $n_\mathrm{inf}$ is also in error relative to the actual average mode occupancy,
\bea
\nbar_{\phi}\arrowvert_{\Delta=\omega_m} - n_\mathrm{inf} = \frac{\kappa_e}{\kappa} n_\phi\label{eqn:ninf_highC},
\eea
where we have assumed again the high cooperativity, deeply sideband resolved limit.  As mentioned above, similar relations for heating and transduction error may be derived for other forms of classical noise, such as laser intensity noise and internal cavity noise.
  
\subsubsection{Effect on sideband asymmetry thermometry}\label{th_sideband_asymmetry_therm}

As indicated pictorially in Fig.~\ref{fig:squashing_anti_squashing} and mathematically in eqn.~(\ref{eq:I_phi_noise_power}), the contribution to the detected photocurrent noise power spectrum due to laser phase noise takes on a different sign for red and blue detuned driving.  In the evanescent coupling geometry considered, the former causes a dip in the phase noise background at the mechanical frequency resulting in noise \textit{squashing}, while the latter leads to a reflection peak in the photodetected noise corresponding noise \textit{anti-squashing}. For the sideband asymmetry thermometry described in Section~\ref{ss:quantum_asymmetry}, this classical noise asymmetry can mask the quantum asymmetry associated with zero-point fluctuations of the mechanical resonator, and thus must be carefully accounted for in such measurements. 

From eqns.~(\ref{eqn:SII}), (\ref{eqn:SII_blueside}), and (\ref{eq:I_phi_noise_power}) we find the inferred mechanical mode populations for probe measurements at detuning $\Delta=\pm \omega_m$,
\bea
n_\text{inf}^\pm = \frac{\gamma_i n_b}{\gamma_\pm} + \frac{|\gamma_{\text{OM}}|}{\gamma_\pm}\frac{(\kappa/2)^2}{(\omega_m \pm \omega_m)^2 + (\kappa/2)^2} \mp \left( \frac{\kappa_e}{\kappa} \right) \frac{\gamma_i \pm |\gamma_\text{OM}/2|}{\gamma_\pm} n_\phi,
\eea
where $n_b$ is the average phonon occupancy of the mechanical resonator in the absence of the probe field.  In the limit of low probe power (low cooperativity, $C_r=|\gamma_{\text{OM}}|/\gamma_i \ll 1$) applicable to measurements performed in Ref.~\cite{Safavi-Naeini2012} this simplifies to,
\bea
n_\text{inf}^\pm\arrowvert_{C_r \ll 1}  = \frac{\gamma_i n_b}{\gamma_\pm} \mp \frac{\kappa_e}{\kappa}\frac{\gamma_i}{\gamma_\pm} n_\phi.
\eea
Note that in this limit, averaging the two detuning measurements of $n_\text{inf}^\pm$ results in a cancellation of the noise squashing for $\Delta=\omega_m$ and the noise anti-squashing for $\Delta=-\omega_m$.  This hints at a method of accurately determining the unperturbed phonon occupation number, i.e. $n_b =  (\gamma_+ n_\text{inf}^+  + \gamma_- n_\text{inf}^-)/2 \gamma_i$, even in the presence of laser phase noise.  

For thermometry based upon the motional sideband asymmetry (Section~\ref{ss:quantum_asymmetry}), however, the effects of classical laser phase noise can not be so easily separated from the quantum noise asymmetry generated by the zero-point fluctuations of the mechanical resonator and the quantum back-action of the vacuum fluctuations of the probe laser. In the low cooperativity regime, the motional sideband asymmetry parameter described in Section~\ref{ss:quantum_asymmetry} is modified to include the effects of probe laser phase noise,
\bea
\eta^{(\phi)} = \frac{1 + 2\kappa_e n_\phi/\kappa}{n_b - \kappa_e n_\phi / \kappa}.\label{eq:asymmetry_phase_noise}
\eea
One way to sort out asymmetry effects related to classical laser phase noise, as is done below in Section~\ref{ss:asymm_measurements}, is to analyze the dependence of the measured asymmetry on the laser probe power.

\section{Experiment}\label{sec:experiments}

\begin{figure}[ht]
\begin{center}
\includegraphics[width=0.55\linewidth]{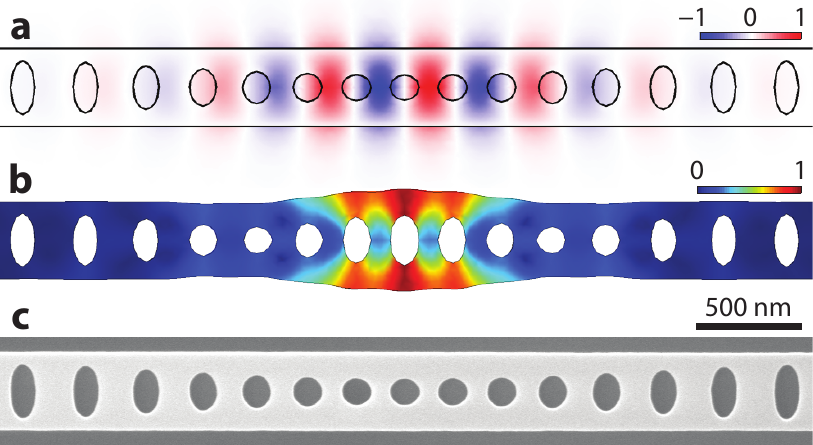}
\caption{\textbf{a}, Finite-element-method (FEM) numerical simulations of the electric field amplitude of the fundamental optical mode (theoretical frequency, $\omega_o/2\pi=194$~THz) of the OMC cavity described in detail in Ref.~\cite{Chan2012}. \textbf{b}, FEM numerical simulation showing the displacement amplitude of the coupled breathing mechanical mode whose theoretical frequency is $\omega_m/2\pi=5.7$~GHz (measured at 5.1 GHz).  \textbf{c}, Scanning electron micrograph of a fabricated version of the silicon nanobeam optomechanical crystal cavity.  
\label{fig:device}}
\end{center}
\end{figure}

Below we describe, using the formalism presented above, several recent experiments~\cite{Chan2011,Safavi-Naeini2012} involving the laser cooling and thermometry of GHz-frequency mechanical resonators formed within optomechanical crystals (OMCs)~\cite{Eichenfield2009b}. Additionally, we present new experimental results, on a more recent OMC cavity~\cite{Chan2012} system, being simultaneously cooled and probed using lasers with and without phase noise.  As depicted in Fig.~\ref{fig:device}, OMCs are engineerable nanoscale structures that may be used to co-localize optical and acoustic (mechanical) waves.  The particular device shown in Fig.~\ref{fig:device}c is a thin ($220$~nm) nanobeam formed from the Si device layer of a silicon-on-insulator wafer which supports a ``breathing mode'' mechanical resonance at a frequency of $5.1$~GHz, and a high-$Q$ optical resonance at an optical frequency of $194$~THz (wavelength $\lambda \approx 1500$~nm).  Theoretical calculation of the optomechanical coupling strength between the breathing mode and co-localized optical mode yields a value of $g_o/2\pi=860$~kHz, due primarily in this design to the elasto-optic effect within Si.  

Three different experiments are described below, each of which is performed with a slightly different OMC device and under slightly different optical coupling conditions.  For reference below, in Table~\ref{tab:devices} we provide the important experimental parameters for all three devices, labelled $A$, $B$, and $C$.  Device $A$ was used in the experiment of Ref.~\cite{Chan2011} to cool a mechanical resonator close to its quantum mechanical ground-state, device $B$ is a newly designed high frequency device~\cite{Chan2012}, and device $C$ was studied in Ref.~\cite{Safavi-Naeini2012} as part of an experiment to measure the effects of quantum zero-point motion.  As indicated, for device $C$, we have both a resonant optical mode used for cooling and a resonant optical mode used for read-out, with both modes coupled to the same mechanical mode.   

\renewcommand{\arraystretch}{1.1}
\renewcommand{\extrarowheight}{0pt}
\begin{table}
\caption{Optomechanical crystal device and measurement parameters}
\label{tab:devices}
\begin{center}
\begin{tabularx}{\linewidth}{YYYYYYY}
\hline
\hline
device & $\lambda_o$ (nm) & $\kappa/2\pi$ (MHz) & $\kappa_e/2\pi$ (MHz) & $g_o/2\pi$ (kHz) & $\omega_m/2\pi$ (GHz) & $\gamma_i/2\pi$ (kHz) \\
\hline 
$A$~\cite{Chan2011} & 1537 & 488 & 65  & 910 & 3.68 & 35 \\
$B$ & 1547.3 & 694 & 97 & 910 & 5.1 & 12.2 \\
$C$~\cite{Safavi-Naeini2012}(cooling) & 1460 & 390 & 46 & 960 & 3.99 & 43 \\
$C$~\cite{Safavi-Naeini2012}(read-out) & 1545 & 1000 & 300 & 430 & 3.99 & 43 \\
\hline
\hline
\end{tabularx}
\end{center}
\end{table}
\renewcommand{\arraystretch}{1.0}
\renewcommand{\extrarowheight}{0pt}

\subsection{Experimental set-up}\label{set_up}

\begin{figure}[ht]
\begin{center}
\includegraphics[width=\linewidth]{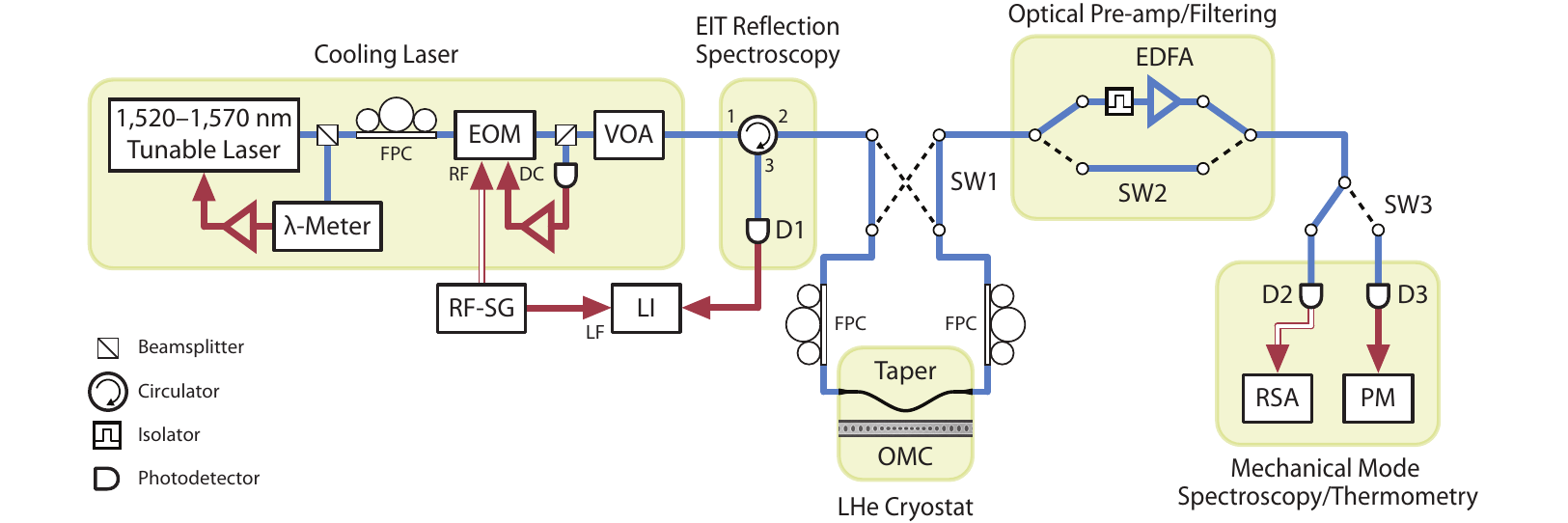}
\caption{\textbf{Laser cooling and calibrated thermometry experimental setup.} As indicated by the separate boxed areas, the experimental setup consists of a tunable laser source used for cooling and mode thermometry, the OMC device under study which is placed in a continous-flow liquid helium cryostat, an optical pre-amplification stage used to amplify the cooling beam signal after transmission through the OMC cavity, a high-speed photodiode and electronic spectrum analyzer for measuring the photocurrent noise power spectral density of the transmitted cooling beam signal, and a modulation/lock-in scheme for probing the near-resonance optical reflection of the OMC cavity in the presence of the strong cooling beam tone (EIT-like reflection spectroscopy). The blue lines indicate the optical path for the cooling and mechanical mode thermometry measurement, while the dashed black lines indicate the alternative switched paths for calibrating the mode thermometry.  The laser source is an external cavity semiconductor diode laser (New Focus Velocity laser, either model TLB-6328 or model TLB-6728), which can have its wavelength tuned over an approximately $60$~nm wavelength span centered around $\lambda=1550$~nm.  Other acronyms are: $2$x$2$ optical switch (SW), variable optical attenunator (VOA), electro-optic modulator (EOM), lock-in detector (LI), erbium-doped fiber amplifier (EDFA), real-time spectrum analyzer (RSA), power meter (PM), optical detector (D), fiber polarization controller (FPC), optical wavemeter ($\lambda$-meter), and RF signal generator (RF-SG).    
\label{fig:exp_setup}}
\end{center}
\end{figure}

The basic experimental set-up used for laser cooling and calibrated thermometry of the optomechanical crystal devices is shown in Fig.~\ref{fig:exp_setup}.  A tunable external cavity semiconductor laser source (New Focus Velocity series) is used both as the cooling laser beam, and upon transmission through the OMC cavity, to perform mode thermometry on the optically coupled mechanical breathing mode of the OMC device\footnote{Measurements of the laser phase noise of each of the different laser models used in the measurements described below are presented in \ref{ss:broadband_measurement}}.  This experimental apparatus allows one to accurately set and measure the laser power at the input of the OMC cavity, and to calibrate the transduced mechanical noise power spectrum that is imprinted on the transmitted cooling laser beam.  The series of micro-electro-mechanical $2$x$2$ optical switches are used to switch the optical path repeatedly into a variety of different configurations, with very little variance ($<1\%$) in power levels.  An optical fiber taper, formed by heating and stretching a single mode SMF28 optical fiber down to a $\sim2$~$\mu$m diameter, is used to evanescently couple light into and out of the OMC device.  This allows the use of fiber optics throughout the set-up for distribution of optical signals, providing a highly stable set-up in which measurments can be performed over days or even weeks.  In order to provide a modicum of pre-cooling, the OMC devices are mounted into a continuous flow liquid helium cryostat, with an attainable temperature of $\sim6$~K measured on the sample mount stage\footnote{The temperature of the breathing mode is  calibrated using the methods described in the main text and is used to extract the temperature  $T_b$ of the mode's locally coupled bath. This bath temperature is typically higher, at $T_b \approx 10-20$~K due to blackbody heating of the sample through the imaging viewport of the cryostat, and imperfect thermalization of the sample surface to the stages.}.  A series of attocube piezo and slip-stick stages are used to position the fiber taper in the near-field of a chosen OMC device on the sample.  Typically the fiber taper is aligned roughly parallel to the nanobeam OMC device, and placed in contact with the surrounding sample surface (but not the nanobeam itself) roughly $100$~nm to one side of the OMC cavity.  A Teflon Swagelok fitting with a pair of small diameter holes drilled in it are used to feed the optical fiber taper into and out of the crystat.

During a laser cooling and thermometry run, a series of measurements are performed at each laser cooling power.  These include: (i) the setting and stabilization of the cooling beam laser frequency to $\Delta=\omega_m$, (ii) calibration of the optical transmission, amplification, and detection, and (iii) measurement of the laser cooled mechanical noise power spectral density and noise background level.  In the first step, the laser cooling beam frequency is set to a mechanical frequency red detuned of the optical cavity resonance ($\Delta=\omega_m$) using near-resonance reflection spectroscopy of the driven cavity system.  Such reflection spectroscopy, what we call EIT-like spectroscopy (see Ref.~\cite{Safavi-Naeini2011}), involves the use of a weak optical sideband of the intense laser cooling beam.  The optical sideband is generated via the EOM, and is swept across a frequency span of $\Delta^{\prime}=1$-$8$~GHz using the RF-SG.  A small amplitude modulation is also applied to the optical sideband at a frequency of $\omega_{\text{LI}}=100$~kHz, allowing for lock-in (LI) detection of the reflected sideband signal from the OMC cavity.  The cooling beam frequency is adjusted until the reflected sideband is aligned with the optical cavity resonance at a modulation frequency equal to the mechanical frequency, $\Delta^{\prime}=\omega_m$\footnote{Due to interference between light which is directly coupled into and out of the cavity, and light which is coupled into the cavity and interacts with the mechanical resonance, the corresponding reflected optical signal of the weak sideband has in addition to the normal broad Lorentzian feature of the optical cavity, a narrowband dip in the reflection.  This reflection dip is analogous the transparency window in atomic EIT, and has a bandwidth of the laser cooled and damped mechanical resonance, with center frequency one mechanical frequency blue-shifted from the laser cooling beam frequency}.  The cooling beam laser frequency is then locked to within $\pm 5$~MHz  of this point using the optical wavemeter ($\lambda$-meter).  

Calibration of the delivered laser cooling beam power to the input of the OMC cavity, and of the photodetected signal of the cooling beam transmission through the OMC cavity, are performed as described in detail in the Supplementary Information to Ref.~\cite{Chan2011}.  In brief this involves measurment of the optical power at important (fiber taper input, fiber taper output, high-speed photoreceiver, etc.) points along the optical signal path shown in Fig.~\ref{fig:exp_setup}.  Measurement of the transmitted power at detector (D3) is used as a reference, being constantly monitored using switch SW3 during the calibration process and throughout subsequent measurements.  Calibration also involves the measurement of the optically induced damping by the laser cooling beam, via measurement of the linewidth of the optically transduced thermal noise power spectrum of the breathing mode, for light sent in both directions through the fiber taper (enabled by the two $2\times2$ optical switches at the input and output of the fiber taper, SW2 and SW3 of Fig.~\ref{fig:exp_setup}).  Asymmetries in the optically induced damping allow one to determine the optical loss before and after the OMC cavity in the fiber taper, which in combination with the total insertion loss of the fiber taper, provides an accurate estimate of the optical input power directly at the cavity.  Calibration of the response of the high-speed photoreceiver (D2) and the optical amplifier (EDFA) is performed before every measurement point by applying an amplitude modulation on the input laser beam (using the EOM of Fig.~\ref{fig:exp_setup}) of known amplitude at a frequency of the mechanical mode of the device under test, and recording the measured response in the photodetected (D2) noise spectrum.

Finally, measurement of the laser-cooled mechanical noise spectrum requires three spectra for proper calibration: the mechanical noise spectrum taken with the cooling laser tuned to the optimal cooling point ($\Delta=\omega_m$), a ‘dark’ spectrum taken with the cooling laser far detuned from the cavity  ($\Delta > 4\omega_m \gg \kappa$), and a background spectrum with the cooling laser blocked.  Comparing the first and second spectra, we obtain the classical noise properties of the laser beam, while the relation between the first and second spectra provide us with the amplifier and detector shot noise levels inherent to the measurement apparatus.

\subsection{Laser cooling and cooling beam mode thermometry}\label{ss:cavity_measurements}

\begin{figure}[ht]
\begin{center}
\includegraphics[width=\linewidth]{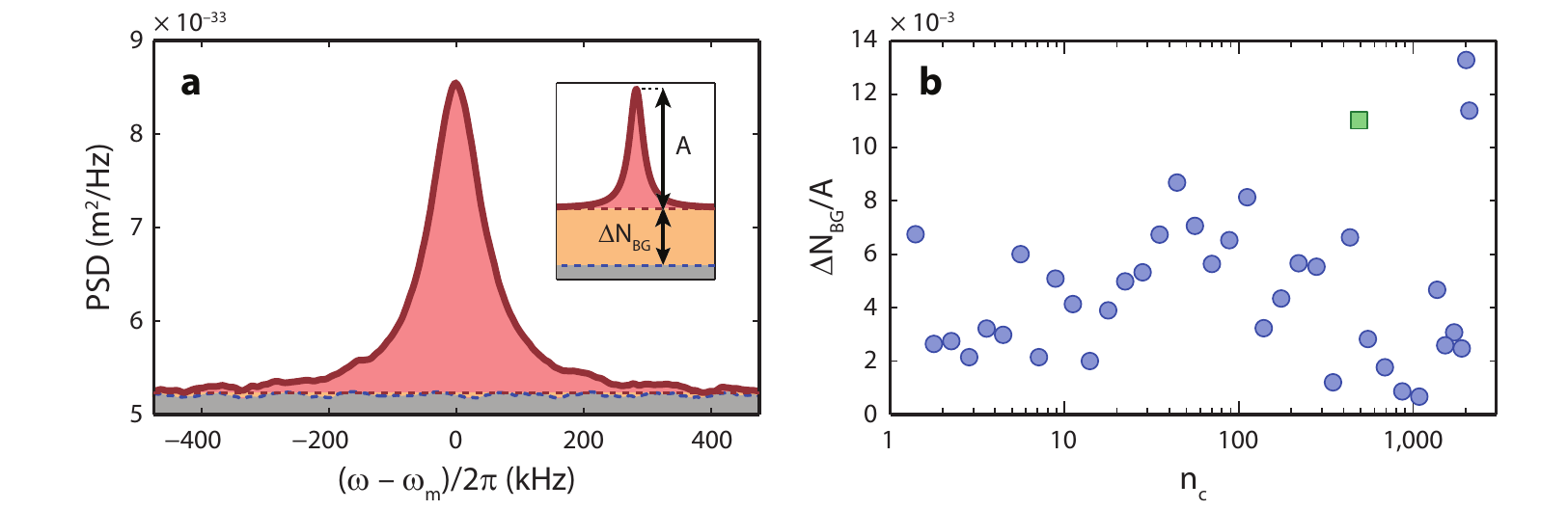}
\end{center}
\caption{\textbf{a}, Measured noise power spectral density near the mechanical resonance frequency of the transmitted optical cooling beam with detuning $\Delta=\omega_m$.  This data corresponds to the cooling point at $m_c=10$ intra-cavity photons of the cooling laser in Fig.~4b of Ref.~\cite{Chan2011}, and appeared in the Supplementary Information of that work.  \textbf{b},  Normalized background noise levels (see text) versus cooling beam intra-cavity photon number, again from the experiment reported in Ref.~\cite{Chan2011}. The green square point ($\square$) was taken with the cooling laser beam filtered through a narrowband ($50$~MHz) optical filter prior to being input to the optomechanical cavity. 
\label{fig:calibrated_background}}
\end{figure}

As described above, the mechanical mode occupancy can be determined by careful calibration of the optically-transduced mechanical noise power spectral density imprinted on the laser cooling beam.  In the Supplementary Information of Ref.~\cite{Chan2011} we describe a method which can be used to both calibrate the laser cooled mechanical mode occupancy as well as to bound the effects due to classical laser noise.  A plot of a typically measured noise spectrum in that experiment is shown in Fig.~\ref{fig:calibrated_background}a, taken under optimal cooling conditions with $\Delta=\omega_m$.  The effect of classical laser noise on heating and thermometry can be discerned by the level of the noise background in the vicinity of the mechanical resonance peak.  One can separate out additional classical laser noise from laser shot noise and detector noise in the measured noise background by making an additional noise measurement with the laser detuned by several mechanica frequencies from the optimal cooling detuning point, $\Delta>4\omega\gg\kappa$.  This changes the measured classical laser noise in the vicinity of the mechanical resonance peak due to the filtering properties of the optical cavity.  In the case of phase noise, as shown in eqn.~(\ref{eqn:SII_phasenoise}), the background noise floor around the mechanical resonance drops by a factor proportional to $1+(\kappa_e/\kappa)^2n_{\phi}$ when the laser detuning shifts from $\Delta=\omega_m$ to $\Delta>4\omega_m$.  Similar relations can be derived in the case of laser intensity noise or intra-cavity noise (such as might result from thermo-refractive noise of the cavity).    

Referring to the inset in Fig.~\ref{fig:calibrated_background}a, by taking the ratio between the difference in the two background noise measurements ($\Delta\text{BG}$) and the noise peak of the phonon signal ($A$), we find for the case of laser phase noise
\bea
\frac{\Delta\text{BG}}{A} = \frac{\kappa_e}{2\kappa}\frac{n_\phi}{n_\text{inf}}\frac{\gamma}{\gamma_\text{OM}} \approx  \frac{\kappa_e}{2\kappa}\frac{n_\phi}{n_\text{inf}},
\eea
in the large cooperativity limit.  Measurements from the ground-state cooling experiment of Ref.~\cite{Chan2011} are shown in Fig.~\ref{fig:calibrated_background}b for the normalized background level ($\Delta\text{BG}/A$). These results show a normalized noise level less than $1.5\%$ for all cooling beam powers (intra-cavity photon numbers).  For the cavity coupling rates of the device measured in Ref.~\cite{Chan2011} (device $A$ in Table~\ref{tab:devices}), this yields a bound on the laser phase noise photon number near mechanical resonance of $n_\phi(\omega_m)\lesssim 0.19$.  From eqn.(\ref{eqn:SII_phasenoise}) and Table~\ref{tab:devices}, for this level of laser noise the mechanical mode heating is $(\kappa_e/2\kappa)n_{\phi}\lesssim 0.012$ quanta and the amount of noise squashing is $(\kappa_e/\kappa)n_{\phi} \lesssim 0.024$ quanta, much less than the smallest inferred phonon occupancy $n_{\text{inf}}=0.85$ of the laser-cooled mode.  As such, laser phase noise heating and squashing in these measurements were determined to be insignificant.  A further verification of this is that no significant change in the noise background level or mechanical mode peak was detected with the cooling laser beam transmitted through an additional scanning Fabry-P\'erot filter (bandwidth $50$~MHz), placed at the input to the optomechanical cavity.  The normalized noise level for this filtered laser measurement is shown alongside the original data presented in Supplementary Information of Ref.~\cite{Chan2011} as a green square point in Fig.~\ref{fig:calibrated_background}b.      

Comparison of the above technique to more conventional laser phase noise measurements presented in \ref{ss:broadband_measurement} can be made by converting the estimated noise quanta into units of laser frequency noise.  For the largest cooling beam intra-cavity photon number used in the experiment of Ref.~\cite{Chan2011} of $n_c\approx2000$ (corresponding to input laser power of $|E_0|^2=8.33\times 10^{14}$ photons/s), the laser phase noise is bounded by (see eqn.~(\ref{eqn:S_EE_Sphiphi})), $\bar{S}_{\phi\phi} = (n_{\phi}/|E_0|^2) \approx 2.3 \times 10^{-16}$~Hz$^{-1}$ at $3.68$~GHz.  For the Model 6328 laser used in this experiment, the corresponding laser frequency noise at $3.68$~GHz is thus at most $\bar{S}_{\omega\omega} = \omega_m^2 \bar{S}_{\phi\phi} \approx 1.2 \times 10^{5}~\text{rad}^2\text{Hz}$, which is consistent with the measured laser frequency noise at this frequency of $\bar{S}_{\omega\omega} \approx 7 \times 10^{4}~\text{rad}^2\text{Hz}$, shown in Fig.~\ref{fig:all_data}a using the calibrated Mach-Zender technique described in ~\ref{ss:broadband_measurement}.  This also explains why there is no obvious trend in the normalized noise level versus laser power of Fig.~\ref{fig:calibrated_background}b; laser phase noise in these measurements is small and is masked by the uncertaintities in the measured background levels.  

\begin{figure}[ht]
\begin{center}
\includegraphics[width=\linewidth]{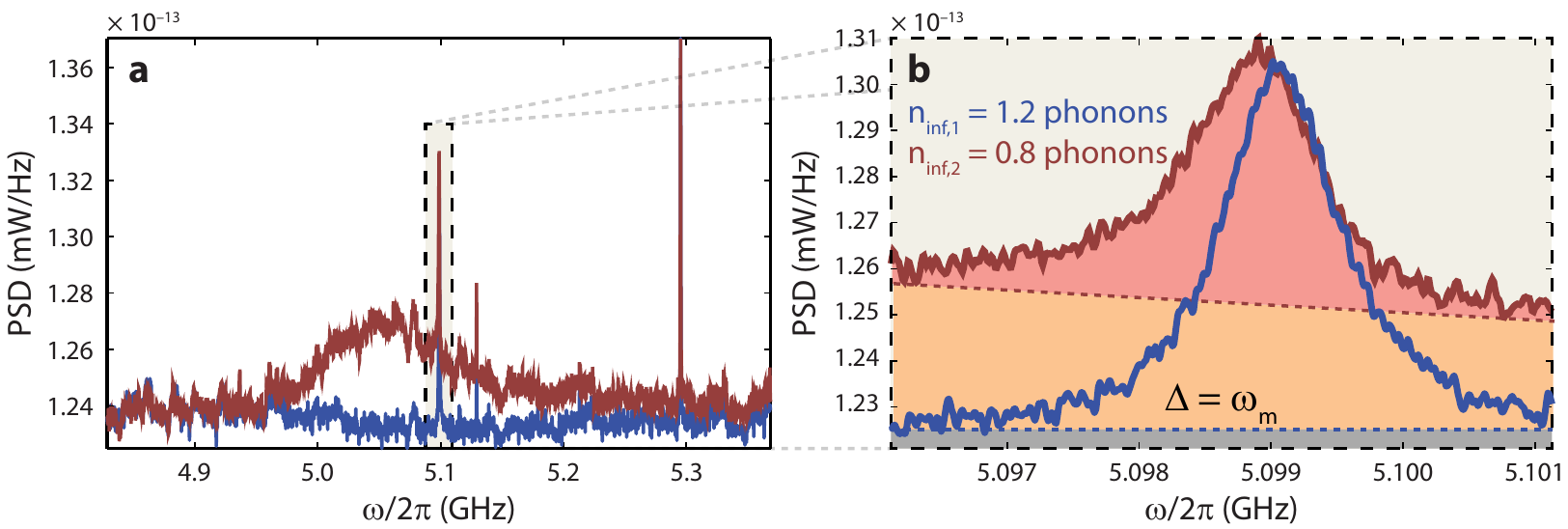}
\end{center}
\caption{\textbf{a} Broadband noise power spectral densities of the high frequency device $B$, measured using two different cooling beam lasers (detuning $\Delta=\omega_m$), one posessing significant phase noise near the mechanical frequency (red curve; Model 6728), the other near quantum-limited (blue curve; Model 6328).  \textbf{b} Zoom-in of the measured noise power spectral density around the cooled phonon peak at $5.1$~GHz.
\label{fig:exp_squashing}}
\end{figure}

In order to actually measure the deleterious effects of laser phase noise on the cooling and mechanical mode thermometry, we have also performed measurements of an OMC device with mechanical resonance at $5.1$~GHz (device $B$ in Table~\ref{tab:devices}).  From phase noise measurements over a wider span ($1$-$7$~GHz) of the Model 6328 and Model 6728 lasers studied in \ref{ss:broadband_measurement}, we found additional phase noise peaks at the second harmonic of the fundamental phase noise peaks shown in Fig.~\ref{fig:all_data}.  For laser wavelengths near the optical resonance of device $B$ ($\lambda=1547.3$~nm), this results in second harmonic phase noise peaks at $5$~GHz and $6$~GHz for the Model 6728 and Model 6328 lasers, respectively.  The phase noise peak at $\omega/2\pi \approx 5~\text{GHz}$ of the Model 6728 laser was measured to have a corresponding frequency noise spectral density of $\bar{S}^{\text{peak}}_{\omega\omega} = 7.5\times 10^6~\text{rad}^2\text{Hz}$, calibrated using the same techniques described in \ref{ss:broadband_measurement}.  The measured phase noise of the Model 6328 laser is roughly $20$~dB smaller than that of the Model 6728 laser at $5$~GHz.  The measurement was set up so that the cooling laser could be switched between the noisy Model 6728 laser and the quiet (in this frequency band) Model 6328 laser used in the previously described ground-state cooling measurements above.  Care was taken to match the laser power and polarization incident on the cavity for both lasers, thus enabling direct comparison of the measured results with differing levels of phase noise present. 

Figure~\ref{fig:exp_squashing} shows the results of such a measurement on device $B$, with red and blue curves corresponding to measurements with the noisy (Model 6728) and quiet (Model 6328) lasers, respectively. Both of these measurements were taken for an optimal cooling detuning of $\Delta/2\pi=\omega_m/2\pi=5.1$~GHz and an input laser power of $|E_0|^2 = 8.1\times 10^{14}$ photons/s (corresonding to an intracavity photon number of $m_c=240$).  The effects of laser phase noise from the Model 6728 laser are immediately evident in the broad frequency scan of Fig.~\ref{fig:exp_squashing}a, in which a broad noise peak can be seen around the mechanical resonance of interest at $5.1$~GHz (there are in fact three narrow mechanical resonance lines visible in the data; the breathing mode mechanical resonance of interest which is most strongly coupled to the optical mode is the one at $5.1$~GHz).  Note that there is no such broad noise peak discernable for the blue curve of the quiet Model 6328 laser.  A zoom-in of the measured mechanical resonance for both lasers is shown in Fig.~\ref{fig:exp_squashing}b.  

A fit to the area under the  mechanical resonance yields an inferred phonon occupancy of $n_{\text{inf}}=0.8$ for the noisy Model 6728 laser and $n_{\text{inf}}=1.2$ for the quiet Model 6328 laser.  $n_{\phi}$ can be estimated from the inferred phonon occupancy and the ratio of the mechanical resonance peak height ($A$) to the laser phase noise background ($N_{\text{BG}}$).  From eqn.~(\ref{eqn:SII_phasenoise}), $n_{\phi} = (N_{\text{BG}}/A)(2\kappa/\kappa_e)n_{\text{inf}}$.  For the red curve in Fig.~\ref{fig:squashing_anti_squashing}b of the Model 6728 laser this yields $n_{\phi}\approx3.2$ noise quanta, corresponding to $(\kappa_e/2\kappa)n_{\phi}\approx 0.4$ phonons of heating and twice that of noise squashing, resulting in an actual mechanical mode occupancy of $\nbar_{\phi}=1.6$ phonons (note that this is larger than the inferred phonon occupancy using the quiet Model 6328 laser by the $0.4$ phase noise heating phonons, as one would expect).  Putting $n_{\phi}$ in units of laser phase noise, we have from eqn.~(\ref{eqn:S_EE_Sphiphi}) that $\bar{S}_{\phi\phi} = (n_{\phi}/|E_0|^2) \approx 4 \times 10^{-15}$~Hz$^{-1}$ at $5.1$~GHz for the Model 6728 laser.  From the broad spectrum in Fig.~\ref{fig:squashing_anti_squashing}a, the peak laser phase noise occurs at $\omega/2\pi=5.05$~GHz, and is approximately $1.65$ times the phase noise at the mechanical resonance frequency, $\bar{S}_{\phi\phi} (\omega/2\pi=5.05$~GHz$) \approx 6.7 \times 10^{-15}$~Hz$^{-1}$.  This corresponds to a peak laser frequency noise of $\bar{S}_{\omega\omega} = \omega^2 \bar{S}_{\phi\phi} \approx 6.8 \times 10^{6}~\text{rad}^2\text{Hz}$, which is in good agreement with the independently calibrated peak laser frequency noise for this laser of $7.5 \times 10^{6}~\text{rad}^2\text{Hz}$.          

\subsection{Sideband asymmetry mode thermometry}\label{ss:asymm_measurements}

In addition to the calibrated mode thermometry measurements described in the experiments above, we have also recently performed a form of self-calibrated measurement of the mechanical mode occupancy involving the measurement of the asymmetry in the motional sidebands generated by laser scattering from an optomechanical crystal resonator near its quantum ground-state~\cite{Safavi-Naeini2012,Khalili2012}.  The device studied in this measurement is device $C$ of Table~\ref{tab:devices}, and it has two optical modes coupled to the same breathing mechanical mode of frequency $\omega_m/2\pi=3.99$~GHz.  One of the optical cavity modes, resonant at wavelength $\lambda_c=1460$~nm, is used to cool the breathing mechanical mode.  The other optical mode, resonant at wavelength $\lambda_r=1545$~nm, is used to read out the mechanical motion via the red (Stokes) and blue (anti-Stokes) scattered sidebands of a read-out laser beam near resonance with the read-out cavity mode.  Further details of the measurement can be found in Ref.~\cite{Safavi-Naeini2012}, along with further discussion of the interpretation of the measurement results in Ref.~\cite{Khalili2012}.  Here we aim to re-present the data of Ref.~\cite{Safavi-Naeini2012}, in a form suitable for analysis of the effects of laser noise (primarily phase noise) as per the theoretical analysis described above in section~\ref{th_sideband_asymmetry_therm}.  We also compare the measured results with the expected noise effects from the calibrated laser phase noise of the read-out laser presented in ~\ref{ss:broadband_measurement}.   

\begin{figure}[ht]
\begin{center}
\includegraphics[width=\linewidth]{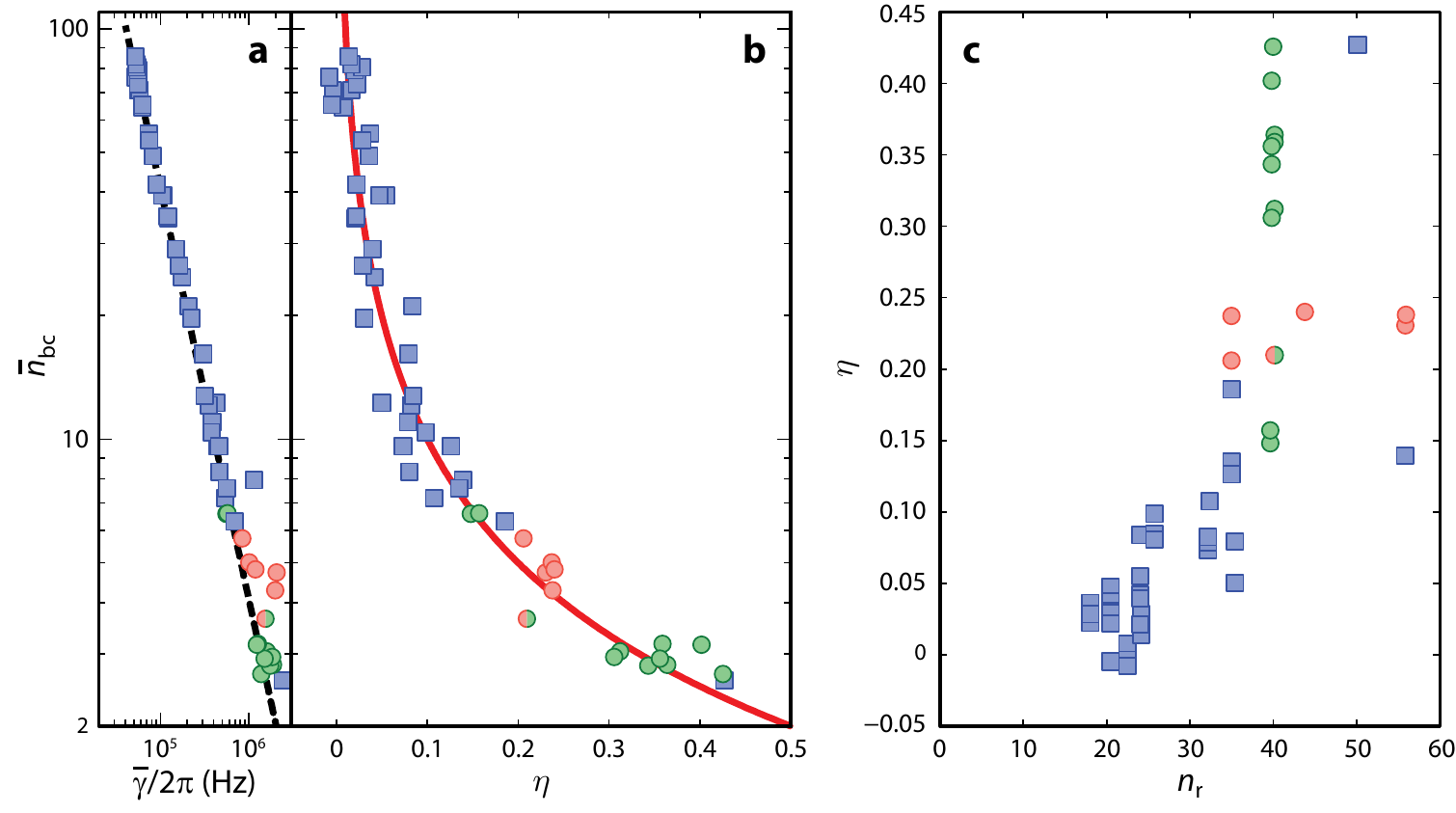}
\end{center}
\caption{ \textbf{a} and \textbf{b} show laser cooling and motional sideband asymmetry data from the experiment reported in Ref.~\cite{Safavi-Naeini2012}. In this experiment a separate cooling beam and probe beam were used, each coupled to the localized breathing mechanical mode of a Si nanobeam resonator at frequency $\omega_m/2\pi = 3.99$~GHz (device $C$ of Table~\ref{tab:devices}).  In \textbf{a} we plot the measured optically enhanced mechanical damping rate ($\bar{\gamma}$) versus the inferred laser cooled mode occupancy of the breathing mechanical resonator ($\bar{n}_{bc}$).  Here $\bar{n}_{bc}$ represents the breathing mode occupancy due to laser cooling in the absence of the probe beam.  It is computed as $\bar{n}_{bc} = (\gamma_+ n^{+}_{\text{inf}} + \gamma_i n^{-}_{\text{inf}})/2\bar{\gamma}$, from the inferred mode occupancies ($n^{\pm}_{\text{inf}}$) and mechanical linewidths ($\gamma_{\pm}$ and $\bar{\gamma} \equiv (\gamma_+ + \gamma_-)/2$) measured with the weak probe beam at detunings $\Delta_r=\pm \omega_m$.  As plotted in \textbf{b}, the weak probe beam signals for detunings $\Delta=\pm \omega_m$ are also used to calculate $\eta$ of eqn.(\ref{eqn:eta}).  The red line in \textbf{b} represents the expected theoretical asymmetry ($\eta=1/\bar{n}_{bc}$) due to zero-point motion of the mechanical system in the absense of phase-noise effects.  In \textbf{c} we also plot the measured asymmetry versus the probe read-out power as represented by the read-out intra-cavity photon number $m_r$.  We have highlighted two subsets of data within the full set of data, represented by green and red circles.  For the green circle data points the measured asymmetry is roughly constant at $\eta=0.23\pm 0.02$ while the probe beam power is varied over a factor of nearly two.  For the red circle data, the probe beam power is held fixed at $n_r=40$ and the asymmetry is measured to range over a factor of three. 
\label{fig:asym_results}}
\end{figure}

Figure~\ref{fig:asym_results}a shows a plot of the inferred phonon occupancy of the breathing mechanical mode at $3.99$~GHz versus the mechanical linewidth of the beathing mechanical mode ($\bar{\gamma}_c$) for each cooling laser beam power.  The cooling laser, a Model 6326 Velocity laser in this case, is tuned to a mechanical frequency red of the fundamental optical mode at $\lambda_c=1460$~nm ($\Delta_c=\omega_m$).  The inferred phonon occupancy in the absence of a read-out beam, $\bar{n}_{bc}$ is measured using a read-out laser (the same Model 6328 Velocity laser used in the ground-state cooling experiments of Ref.~\cite{Chan2011}) that is tuned near resonance of the second-order optical mode of the OMC cavity at $\lambda_r=1545$~nm.  This occupancy ($\bar{n}_{bc}$) can be thought of as the occupation number of a new effective thermal bath coupled to the mechanics consisting of a combination of the intrinsic mechanical damping and optical damping from the cooling beam. $\bar{n}_{bc}$ is measured by taking the average of the calibrated phonon occupancies measured for read-out beam detunings of $\Delta_r=\pm\omega_m$; $\bar{n}_{bc}\equiv 1/2(\gamma_{+}\bar{n}_{+} +\gamma_{-}\bar{n}_{-})/\bar{\gamma}$, where $\bar{n}_{+}$ and $\bar{n}_{-}$ are the inferred phonon occupancies from the measured read-out beam photocurrent spectrum and $\gamma_{+}$ and $\gamma_{-}$ are the measured mechanical linewidths for $\Delta_r=\omega_m$ and $\Delta_r=-\omega_m$, respectively.  $\bar{\gamma}$ is the mechanical damping rate of the total system in the absence of the a read-out beam and is calculated from the measured probe beam spectra using $\bar{\gamma}\equiv 1/2(\gamma_{+} + \gamma_{-})$.  At each of the cooling beam powers, the read-out laser beam power is adjusted such that the cooperativity of the read-out beam is substantially below unity (see Ref.~\cite{Safavi-Naeini2012}).  As such, using this two-mode approach, one high power for cooling and one low power for read-out, separates the noise heating and noise squashing effects, and substantially reduces any effects due to laser phase noise on the mode thermometry.     

In Fig.~\ref{fig:asym_results}b we plot the measured asymmetry in the motional sidebands, $\eta$ in eqn.~(\ref{eqn:eta}), of the read-out laser beam versus the inferred laser-cooled mode occupancy of Fig.~\ref{fig:asym_results}a.  The red solid curve of Fig.~\ref{fig:asym_results}b also plots the expected sideband asymmetry from the linear fit of the laser-cooled mode occupancy versus mechanical damping (dashed black-curve of Fig.~\ref{fig:asym_results}a).  The asymmetry is quite pronounced at lower mode occupancies as expected, and matches well the theoretical sideband asymmetry curve of eqn.~(\ref{eqn:eta}) due to the behavior of the zero-point fluctations of the mechanical mode and interference with the quantum back-action of the read-out laser shot-noise~\cite{Safavi-Naeini2012,Khalili2012}.  As discussed in Section~\ref{th_sideband_asymmetry_therm}, such sideband asymmetry can also arise from classical laser noise on the red-out laser beam (see eqn.(\ref{eq:asymmetry_phase_noise})).  Owing to the fact that the read-out laser beam is of substantially lower power than that of the cooling laser beam in these experiments, the amount of classical laser phase noise is expected to be small.  Also, the read-out laser beam power was only weakly correlated with the cooling beam power (and thus laser-cooled mode occupancy), making it highly unlikely that the measured motional sideband asymmetry curve of Fig.~\ref{fig:asym_results}b is due to classical laser noise on the read-out beam.  Verification of this is shown in Fig.~\ref{fig:asym_results}c, in which a scatter plot of the measured sideband asymmetry is plotted versus the read-out beam laser power (intra-cavity photon  number, $n_r$).  We have highlighted in this set of data two groups of points.  The red circles  correspond to a range of data points in which the read-out beam power is varied over a factor of nearly two ($n_r\approx 30$-$60$), whereas the measured asymmetry varies by less than $\pm 10\%$ of its nominal value ($\eta=0.23\pm0.02$).  The green circle points show data in which the read-out beam power was held fixed ($n_r\approx 40$) as the cooling beam power was varied.  For these data points the measured asymmetry is seen to vary by a factor of almost three ($\eta\approx 0.15$-$0.45$).  Both of these groups of data points serve to indicate a lack of correlation between the measured motional sideband asymmetry and the read-out beam power, strongly ruling out laser phsae noise as the source of the measured asymmetry in the experiments of Ref.~\cite{Safavi-Naeini2012}.                

Finally, one can estimate the magnitude of the effects of laser phase noise of the read-out laser on the measured sideband asymmetry in these measurements by comparing to the calibrated laser phase noise presented in \ref{ss:broadband_measurement}.  From Fig.~\ref{fig:all_data}a, the laser frequency noise of the Model 6328 laser used in the read-out of the sideband asymmetry experiment has a value of $\bar{S}_{\omega\omega} \approx 5\times 10^4~\text{rad}^2\text{Hz}$ at a frequency of $\omega/2\pi=3.99$~GHz (the breathing mechanical mode frequency of device $C$) and for a laser wavelength near $\lambda_r=1545$~nm.  From Fig.~\ref{fig:asym_results}c, the read-out beam laser power is at most $|E_0|^2_r = n_r/(\kappa_e/2\omega_m^2) \lesssim 4\times 10^{13}$~photons/s.  The maximum read-out laser phase noise quanta in these measurements is then bounded by $n_{\phi} = (\bar{S}_{\omega\omega}/\omega_m^2)|E_0|^2_r \lesssim 0.003$ quanta.  The corresponding laser noise heating of the mechanical resonator by the weak read-out beam is $(C_r/1+C_r)(\kappa_e/2\kappa)n_{\phi}$, which for the read-out beam cooperativity of these experiments ($C_r \lesssim 0.1$) is at most a negligible $4.3 \times 10^{-5}$ phonons.  Laser phase noise squashing and anti-squashing results in the modified motional sideband asymmetry given in eqn.~(\ref{eqn:quantum_phasenoise_limit_damping}) for a low-cooperativity read-out beam.  The correction factor to the quantum asymmetry due to classical laser phase noise is given by $(1+2\kappa_e n_{\phi}/\kappa)/(1-\kappa_e n_{\phi}/\kappa\bar{n}_c)-1$, which for the measurement of Ref.~\cite{Safavi-Naeini2012} is smaller than $0.2\%$.  We should also note that from the measured phase noise of the Model 6326 laser used to cool the mechanical mode in these experiments (see Fig.~\ref{fig:all_data}c,f), the estimated laser phase noise heating from the cooling beam is less than $\bar{n}_{\phi} \approx 0.04$~phonons at the largest cooling beam powers ($n_c=330$ intra-cavity photons), indicating that the minimum measured phonon occupancy of $2.6$ is also not limited by laser noise in this experiment. 

\section{Conclusions}

We have presented a unified analysis of the effects of optical noise, due to both laser phase noise and quantum fluctuations of the electromagnetic field, on thermometry and cooling in recent optomechanical experiments. By doing so, we provide a systematic means by which the presence of classical laser noise can be detected, and taken into account, and we further rule out any role which it may play in our recent experiments~\cite{Chan2011,Safavi-Naeini2012}. This result is particularly useful for experimenters working on gigahertz nanomechanical resonators, since it allows, with some care, for investigations to probe qauntum optomechanics in the telecom wavelength bands with cheaper, and more readily commercially available diode lasers.

\section{Acknowledgements}

This work was supported by the DARPA/MTO ORCHID program, the Institute for Quantum Information and Matter, an NSF Physics Frontiers Center with support of the Gordon and Betty Moore Foundation, and the Kavli Nanoscience Institute at Caltech. JC and ASN gratefully acknowledge support from NSERC. SG acknowledges support from the European Commission through a Marie Curie fellowship. HM and YC have also been supported by NSF grants PHY-0555406, PHY-0956189, PHY-1068881, as well as the David and Barbara Groce startup fund at Caltech.

\appendix

\section{Definitions}

Fourier Transforms are defined for operators and variables in the symmetric manner
\begin{eqnarray}
\op{A}{}(t) &=& \frac{1}{\sqrt{2\pi}} \int_{-\infty}^{\infty} d\omega~ e^{-i\omega t}\op{A}{}(\omega), \nonumber\\
\op{A}{}(\omega) &=& \frac{1}{\sqrt{2\pi}} \int_{-\infty}^{\infty} dt~ e^{i\omega t}\op{A}{}(t).
\end{eqnarray}
Spectral densities are defined as
\begin{eqnarray}
S_{AA}(\omega) = \int_{-\infty}^{\infty} d\tau~ e^{i\omega \tau} \avg{\opdagger{A}{}(t+\tau)\op{A}{}(t)}.
\end{eqnarray}
and symmetrized as $\bar{S}_{AA}(\omega) = \frac{1}{2} (S_{AA}(\omega) + S_{AA}(-\omega))$. Here the angular brackets denote expectation values as defined in quantum mechanics  $\avg{A} = \text{Tr}(A \rho)$. When classical stochastic processes are placed in angular brackets, we refer to a classical ensemble average. In Fourier domain, the various system operators are written in terms of the bath noise operators, and therefore knowledge of the expectation values of form $\avg{\op{A}{}(\omega)}$ and $\avg{\op{B}{}(\omega)\op{A}{}(\omega^\prime)}$ where $\op{A}{}$ and $\op{B}{}$ are bath field operators is sufficient to calculate spectral densities. These correlations are known from the density matrix of the baths, which in this paper are assumed to be in either  a vacuum or thermal state.
The Hermitian conjugate of operator $\op{A}{}(t)$ is $\opdagger{A}{}(t)$, and has a Fourier transform denoted as 
$$\opdagger{A}{}(\omega) = \frac{1}{\sqrt{2\pi}} \int_{-\infty}^{\infty} dt~ e^{i\omega t}\opdagger{A}{}(t),$$
leading to $\left(\op{A}{}(\omega)\right)^\dagger = \opdagger{A}{}(-\omega).$ The spectral density may be written also as
\begin{eqnarray}
S_{AA}(\omega) = \int_{-\infty}^{\infty} d\omega^\prime~ \avg{\opdagger{A}{}(\omega)\op{A}{}(\omega^\prime)}\label{eqn:SAA_omega_integral}.
\end{eqnarray}

\section{Mechanical resonator spectral density}\label{app:mech_osc}

We use mainly the input-output formalism to derive the various spectra which are being measured. For the case of
the mechanical resonator, the quantum Langevin equation is given by
\be
\label{eqn:mechlang}
\dot{\op{b}{}}(t) = -\left(i\omega_m+\frac{\gamma_i}{2}\right)\op{b}{}(t) - \sqrt{\gamma_i}\op{b}{\mathrm{in}}(t),
\ee
With correlation functions for the input noise 
\bea
\avg{\opdagger{b}{\mathrm{in}}(t)\op{b}{\mathrm{in}}(t^\prime)} &=& n_b\delta(t-t^\prime),\\
\avg{\op{b}{\mathrm{in}}(t)\opdagger{b}{\mathrm{in}}(t^\prime)} &=& (n_b+1)\delta(t-t^\prime),
\eea
where $n_b$ is the occupancy of the thermal bath connected to the mechanical resonator.  Inherent in the correlation functions above is the assumption that the mechanical resonance bandwidth is very small compared to its resonance frequency (high mechanical $Q$-factor), such that the bath occupation can be taken as a single number $n_b(\omega_m)$. 
Additionally, we find for the Fourier transform of the input noise operators,
\bea
\avg{\opdagger{b}{\mathrm{in}}(\omega)\op{b}{\mathrm{in}}(\omega^\prime)} &=& n_b\delta(\omega+\omega^\prime),\\
\avg{\op{b}{\mathrm{in}}(\omega)\opdagger{b}{\mathrm{in}}(\omega^\prime)} &=& (n_b+1)\delta(\omega+\omega^\prime).
\eea
By solving the Fourier transform of equation~(\ref{eqn:mechlang}), we find that the mechanical mode annihilation operator will be deterimined by the input noise as
\be
\op{b}{}(\omega) = \frac{-\sqrt{\gamma_i}\op{b}{\mathrm{in}}(\omega)}{i(\omega_m - \omega) + \gamma_i/2}.
\ee 
This can easily be used to calculate the spectral density, and we find, 
\bea
S_{bb}(\omega) &=& \int_{-\infty}^{\infty} d\tau~ e^{i\omega \tau} \avg{\opdagger{b}{}(\tau)\op{b}{}}\nonumber\\
&=&  \int_{-\infty}^{\infty} d\tau~   e^{i\omega \tau} \frac{1}{2\pi} \int_{-\infty}^{\infty} d\omega^{\prime\prime}~ \int_{-\infty}^{\infty} d\omega^{\prime}~\nonumber\\
&& ~~~~~~~~~~\times\avg{(\op{b}{}(\omega^{\prime\prime}))^\dagger\op{b}{}(\omega^\prime)} e^{i\omega^{\prime\prime}\tau}\nonumber\\
&=&\frac{\gamma_i \bar{n}}{(\omega_m + \omega)^2 + (\gamma_i/2)^2} \label{eqn:sbb}.
\eea
We will sometimes denote this function as $S_{bb}(\omega;\bar{n})$. This spectra density can be thought to represent the ability of the mechanical system to \textit{emit} energy. A similar expression can be found for the creation operators:
\bea
S_{b^\dagger b^\dagger}(\omega) &=& \frac{\gamma_i (\bar{n}+1)}{(\omega_m - \omega)^2 + (\gamma_i/2)^2} \label{eqn:sbdbd}
\eea
This spectra density can be thought to represent the ability of the mechanical system to \textit{absorb} energy. Finally, for the position operator, $\op{x}{} = x_{\mathrm{zpf}} (\op{b}{} +\opdagger{b}{})$, we find 
\cite{Clerk2010}
\bea
S_{xx}(\omega) &=&  x^2_{\mathrm{zpf}}\left( S_{bb}(\omega) + S_{b^\dagger b^\dagger}(\omega) \right) \nonumber\\
 &=&  x^2_{\mathrm{zpf}}\left( \frac{\gamma_i \bar{n}}{(\omega_m+\omega)^2+(\gamma_i/2)^2} + \frac{\gamma_i (\bar{n}+1)}{(\omega_m-\omega)^2+(\gamma_i/2)^2} \right). \nonumber
\eea

\section{Quantum noise squashing}\label{app:quantum_noise_squashing}

Assuming perfect detection ($\kappa_e/2 = \kappa$) we find that the normalized heterodyne current is given by 
\be
\op{I}{}(t) = - i \op{a}{\mathrm{in}}(t) + i \opdagger{a}{\mathrm{in}}(t) + \frac{2G}{\sqrt{\kappa}} (\op{b}{}(t) +  \opdagger{b}{}(t))
\ee
Taking the autocorrelation of the detected current,
\bea
\avg{\op{I}{}(\tau)\op{I}{}} &=& \avg{\op{a}{\mathrm{in}}(\tau)\opdagger{a}{\mathrm{in}}} + \frac{4|G|^2}{\kappa} \frac{\avg{\op{x}{}(\tau)\op{x}{}}}{x^2_\mathrm{zpf}} - \nonumber\\&&
\frac{2iG}{\kappa} \left( \avg{\op{a}{\mathrm{in}}(\tau)\opdagger{b}{}}  - \avg{\op{b}{}(\tau)\opdagger{a}{\mathrm{in}}} \right)
\eea
we find $S_{II}(\omega)$ by taking the Fourier transform of the above expression. The first term, due to the fact that the noise is delta correlated gives a constant noise floor. The second term can be thought of as a measurement of position, and we see that the rate at which information is gathered about the system is $4|G|^2/\kappa$, i.e. the optomechanical damping rate, and back-action. The cross-correlation terms are calculated as such:
\bea
 \int_{-\infty}^{\infty} d\tau~ e^{i\omega \tau}  \avg{\op{a}{\mathrm{in}}(\tau)\opdagger{b}{}} = \nonumber~~~~~~~~~~~~~~~~~~~~~~~~~~~~~~~~~~~~~~~~~~~~\\
  \frac{1}{2 \pi}\int_{-\infty}^{\infty} d\tau~ e^{i\omega \tau}  \int_{-\infty}^{\infty} d\omega^\prime~ \int_{-\infty}^{\infty} d\omega^{\prime\prime}~e^{-i\omega^\prime \tau}  \avg{\op{a}{\mathrm{in}}(\omega^\prime)(\op{b}{}(\omega^{\prime\prime}))^\dagger} 
\eea
Using the back-action modified mechanical fluctuation operator shown in equation  (\ref{eqn:b_inputs}), 
\bea
\op{b}{}(\omega) &=&  \frac{-\sqrt{\gamma_i}\op{b}{\mathrm{in}}(\omega)}{i(\omega_m-\omega) + \gamma/2} + \frac{2iG}{\sqrt{\kappa}} \frac{\op{a}{\mathrm{in}}(\omega)}{i(\omega_m-\omega) + \gamma/2}
\eea
and the properties of vacuum fluctuation operators, we find
\bea
 \int_{-\infty}^{\infty} d\tau~ e^{i\omega \tau}  \avg{\op{a}{\mathrm{in}}(\tau)\opdagger{b}{}} &=& 
-\frac{2iG}{\sqrt{\kappa}} \frac{1}{-i(\omega_m - \omega) + \gamma/2},\\
 \int_{-\infty}^{\infty} d\tau~ e^{i\omega \tau}  \avg{\op{b}{}(\tau)\opdagger{a}{\mathrm{in}}}&=&
+\frac{2iG}{\sqrt{\kappa}} \frac{1}{i(\omega_m - \omega) + \gamma/2}.
\eea
From here we calculate the spectral density of the heterodyne signal,
\bea
S_{II}(\omega) &=& 1 + \nonumber\\
&& +\frac{4|G|^2}{\kappa} \left( \frac{\gamma \bar{n}}{(\omega_m+\omega)^2+(\gamma/2)^2} + \frac{\gamma (\bar{n}+1)}{(\omega_m-\omega)^2+(\gamma/2)^2} \right) \nonumber\\ 
&&-\frac{4|G|^2}{\kappa} \frac{\gamma}{(\omega_m-\omega)^2+(\gamma/2)^2}\nonumber\\
&=& 1 + \frac{8|G|^2}{\kappa}  \bar{S}_{bb}(\omega).\nonumber
\eea

\section{Scattering matrix elements}\label{app:scattering_matrix}

By algebraic manipulation of the Heisenberg equations of motion in Fourier domain, and an additional input-output boundary condition $\op{a}{\text{out}} = \op{a}{\text{in}} + \sqrt{\kappa_e/2} \op{a}{}$, we arrive at scattering relations
\bea
\op{a}{\mathrm{out}}(\omega)\arrowvert_{\Delta=-\omega_m} \approx t(\omega;\Delta)\op{a}{\mathrm{in}}(\omega) + n_{\mathrm{opt}}(\omega;\Delta)\op{a}{\mathrm{in,i}}(\omega) + s_{12}(\omega;\Delta)\opdagger{b}{\mathrm{in}}(\omega)
\eea
and
\bea
&\op{a}{\mathrm{out}}(\omega)\arrowvert_{\Delta=\omega_m} \approx t(\omega;\Delta)\op{a}{\mathrm{in}}(\omega) + n_{\mathrm{opt}}(\omega;\Delta)\op{a}{\mathrm{in,i}}(\omega) + s_{12}(\omega;\Delta)\op{b}{\mathrm{in}}(\omega),
\eea
for blue- and red-side laser pumping, respectively. In the driven weak coupling regime ($\gamma_\text{OM} \ll \kappa$), these scattering coefficients have simple algebaric forms which are presented below.

\subsection{Red-side driving: $\Delta = \omega_m$}
The red-side scattering matrix elements for values of $\omega$ about the mechanical frequency ($\omega-\omega_m \ll \kappa$) are:
\bea
t(\omega;\Delta=\omega_m) = 1 - \frac{\kappa_e}{\kappa} + \frac{|\gamma_\text{OM}|\kappa_e}{2\kappa} \frac{1}{i(\omega_m-\omega) + \gamma/2},
\eea
\bea
n_\text{opt}(\omega;\Delta=\omega_m) = \sqrt{\frac{2 \kappa^\prime \kappa_e}{ \kappa^2}} \left(\frac{|\gamma_\text{OM}|/2}{i(\omega_m-\omega)+\gamma/2} - 1 \right),
\eea
and
\bea
s_{12}(\omega;\Delta=\omega_m) = \sqrt{\frac{\kappa_e}{2\kappa}} \frac{i\sqrt{\gamma_i |\gamma_\text{OM}|}}{i(\omega_m-\omega)+\gamma/2}
\eea

\subsection{Blue-side driving: $\Delta = -\omega_m$}
The blue-side scattering matrix elements for values of $\omega$ about the mechanical frequency ($\omega+\omega_m \ll \kappa$) are:
\bea
t(\omega;\Delta=-\omega_m) = 1 - \frac{\kappa_e}{\kappa} - \frac{|\gamma_\text{OM}|\kappa_e}{2\kappa} \frac{1}{-i(\omega_m+\omega) + \gamma/2}
\eea
\bea
n_{\mathrm{opt}}(\omega;\Delta=-\omega_m) = -\sqrt{\frac{2\kappa^\prime \kappa_e}{\kappa^2}} \left(\frac{ |\gamma_\text{OM}|/2}{-i(\omega_m+\omega)+\gamma/2}  + 1 \right)
\eea
\bea
s_{12}(\omega;\Delta=-\omega_m) = \sqrt{\frac{\kappa_e}{2\kappa}} \frac{i\sqrt{\gamma_i |\gamma_\text{OM}|}}{-i(\omega_m+\omega)+\gamma/2}
\eea

\section{Laser phase noise measurement}\label{ss:broadband_measurement}

The constraints of silicon nanofabricated devices, i.e. their single-mode nature, and their spread in parameters such as optical cavity frequency caused by fabrication imperfections, mean that wideband tunable external cavity diode lasers (ECDL) are invaluable for experiments in optomechanics. As such, they have been used  extensively by our group as well as others~\cite{Gavartin2011,Li2008,Srinivasan2011}, for a variety of systems operating at different mechanical frequencies. Phase and frequency noise have been of concern in these types of lasers, and it is therefore important to evaluate the laser noise properties.

\begin{figure}[ht!]
\begin{center}
\includegraphics[width=\linewidth]{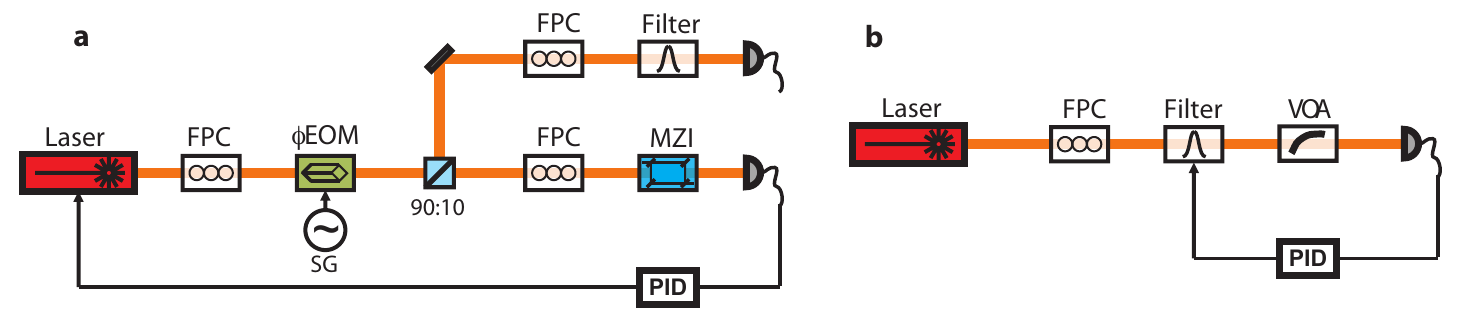}
\end{center}
\caption{\label{fig:setup} \textbf{a} The experimental setup used for measurement and calibration of laser phase noise. Light from the laser under test
is sent into an electro-optical phase modulator ($\phi$EOM), generating a pure
phase modulation which is used to calibrate the system response. Most of
the light is then taken from the 90\% end of a beam splitter, and sent 
through an all-fiber imbalanced ($\Delta L \sim 2$~cm) Mach-Zehnder interferometer (MZI), with a free spectral
range of 20.1~GHz (measured using calibrated Toptica wavemeter). The MZI converts frequency
fluctuations into intensity fluctuations, which are detected on a high speed photodetector (New Focus 1544-B). The DC output of the photodetector is used to perform a low-frequency lock of the laser frequency to the mid-point of the MZI sinusoidal transfer function. The 10\% split-off signal is sent through a scannable optical filter (bandwidth 50MHz), which allows one to independently measure the modulation index of the phase modulated calibration tone by sweeping the filter across the carrier and sidebands, comparing their optical power.  Fiber polarization controllers (FPCs) are used to adjust the optical field polarization. \textbf{b} For an alternate calibration of the laser
phase noise, the shot noise level of the detected laser signal is measured. This is accomplished by sending the
laser through a narrow optical filter (50 MHz bandwidth), followed by a 
variable optical attenuator (VOA), to obtain the shot noise level of the photodetected signal at GHz frequencies. The optical filter is used to remove classical laser noise at these frequencies.}
\end{figure}

Here we present a measurement of $S_{EE}(\omega)$ based on a measurement of the phase noise spectral density $S_{\phi\phi}(\omega)$ and relation (\ref{eqn:S_EE_Sphiphi}). A Mach-Zehnder intereferomoter (MZI) is an optical component where the detected intensity of the transmitted light is dependent on the frequency of the light, and given by $T(\omega) = 1 + \sin(\omega/\omega_\text{FSR})$, where $\omega_\text{FSR}$ is the free spectral range of the interferometer. For the moment, we linearize this relation as, $T(\omega) = 1 + \omega/\omega_\text{FSR}$ (the full relation is used for the presented data). Assuming that the laser frequency $\omega_L = \omega_L(t)$, is a stochastic process representing the instantaneous frequency of the laser light, the detected intensity will be given by $I(t) = \dot{N} + \dot{N} \omega_L(t) / \omega_\text{FSR} + n_\text{SN}(t)$. The term $\dot{N}$ is the average flux of photons incident on the detector, while the last term $n_\text{SN}(t)$, is the white shot-noise of the laser with amplitude proportional to $\dot{N}^{1/2}$. The spectral density of the detected signal is then given by,
\bea
S_{II}(\omega) = \dot{N} + \frac{\dot{N}^2}{\omega^2_\text{FSR}} \bar{S}_{\omega\omega}(\omega).
\eea
Therefore, using the shot noise as a reference level, we can define a signal-to-noise ratio as,
\bea
\text{SNR} =\frac{\dot{N}}{\omega_\text{FSR}^2} \bar{S}_{\omega\omega}(\omega),
\eea
where the ``signal'' is the classical laser frequency noise and the ``noise'' is shot-noise.  This shot-noise based calibration method can be performed by a set-up such as the one shown in Figure \ref{fig:setup}b. Such measurement of shot-noise however depends on knowledge of the detector quantum efficiency.

An alternate calibration method, free of any need to measure absolute powers and detector efficiencies, or to model the optical component transducing the phase noise (here the MZI), is shown schematically in Figure \ref{fig:setup}a. A phase modulator is used to generate a tone with a large modulation index $\beta$ at the frequency of interest. This unitless modulation index is directly obtained by measuring the power of the generated sidebands, using a scanning Fabry-P\'erot filter in our case. These powers are related to each other by the appropriate Bessel function with argument $\beta$. As described below in~\ref{app:calibration}, measuring the tone passing through the MZI provides us with a calibration of $\bar{S}_{\omega\omega}(\omega)$. 

\begin{figure*}[ht!]
\begin{center}
\includegraphics[width=\textwidth]{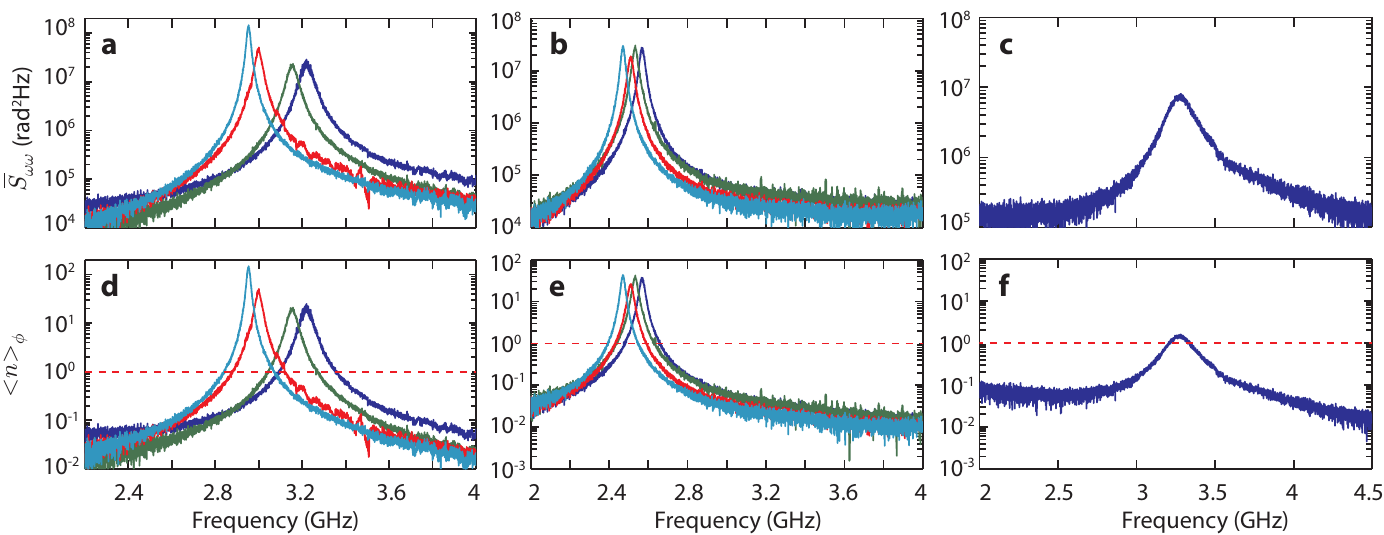}
\end{center}
\caption{\label{fig:all_data} \textbf{a, b} and \textbf{c} are the calibrated
frequency noise $\bar{S}_{\omega\omega}$ for Lasers 1 (Model 6328, Serial \#286), 2 (Model 6728), and 3 (Model 6326, Serial \#19),  
respectively. In \textbf{a, b} the different color curves are for measurements
made at different laser wavelenghts, with \{blue, green, red, cyan\} corresponding to $\lambda=$\{1520, 1537, 1550, 1570\}nm, respectively. Laser
3 is a 1400nm band laser which was only operated at $\lambda=1460$~nm. The plots
shown in \textbf{d, e} show the corresponding level of laser phase noise heating ($\bar{n}_{\phi}$) in units of phonons for the device parameters and maximum cooling laser power ($n_c=2000$ intra-cavity cooling beam photons, $\kappa_e/2\pi=65$~MHz, $\kappa/2\pi=488$~MHz, $\omega_m/2\pi=3.68$~GHz) of Ref. \cite{Chan2011}.  In \textbf{f} we show the same for the device parameters and maximum cooling beam power ($n_c=330$, $\kappa_e/2\pi=45$~MHz, $\kappa/2\pi=300$~MHz, $\omega_m/2\pi=3.99$~GHz) of  Ref. \cite{Safavi-Naeini2012}.}
\end{figure*}

\begin{figure*}[ht!]
\begin{center}
\includegraphics[width=\textwidth]{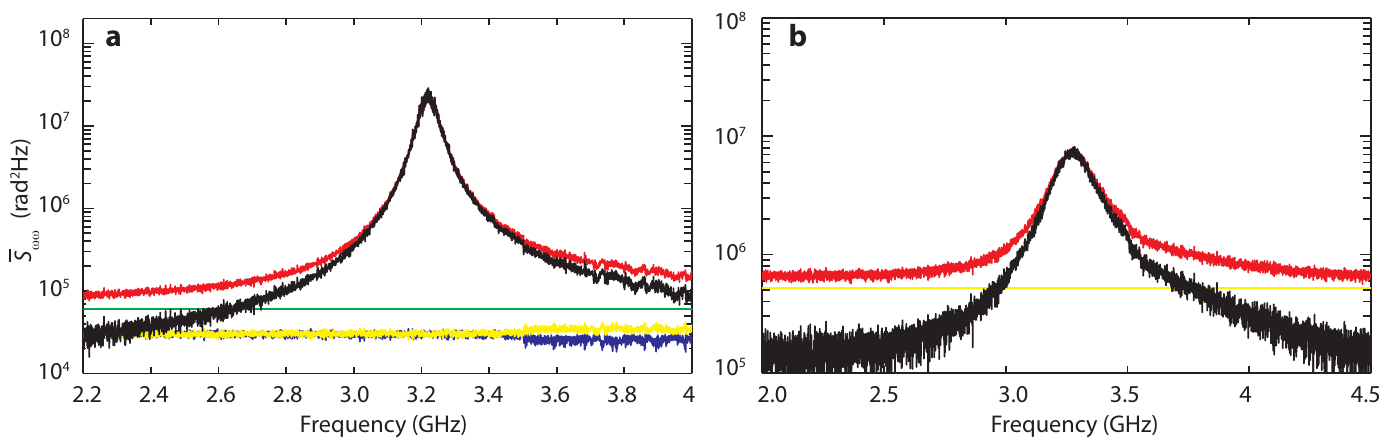}
\end{center}
\caption{\label{fig:all_calibration_data} \textbf{a}, Measured total noise spectrum (red curve), along with the shot noise (blue curve) and detector NEP (yellow curve), used to generate the laser frequency noise spectrum of Fig. \ref{fig:all_data}a (only $1520$~nm wavelength shown).  The detector NEP is measured by simply blocking the laser (zero optical power) and measuring the noise spectrum (note that the detector NEP is well above the intrinsic noise level of the spectrum analyzer).  The shot-noise is measured independently for an identical laser power and wavelength using the set-up of Fig.~\ref{fig:setup}b.  The green line corresponding to the sum of average NEP and shot noise levels.  The black curve is the measured signal with the shot noise and NEP background levels subtracted, and represents the component of the measured noise which we attribute to laser frequency noise in Fig. \ref{fig:all_data}a.  \textbf{b}, Measured total noise spectrum (red curve) and detector NEP (yellow curve) used to generate the laser frequency noise spectrum of Fig. \ref{fig:all_data}c.  The shot-noise level (not shown) for this measurement is far below the NEP level.  The black curve is the measured signal with the NEP subtracted, and is the resulting laser frequency noise spectrum plotted in Fig. \ref{fig:all_data}c.}
\end{figure*}

Both methods give nominally the same results for the calibration of laser frequency noise, which we plot in Fig.~\ref{fig:all_data}a-c for three different New Focus Velocity series external-cavity semiconductor diode lasers used in the work presented here and in recent cavity-optomechanical experiments with GHz-frequency OMCs~\cite{Chan2011,Safavi-Naeini2012}.  In these plots, the detector noise at zero input power (typically denoted as noise-equivalent power (NEP)) and the laser shot noise, are subtracted from raw measured noise spectra. Spectra showing the raw data at various stages of the processing are shown in Fig.~\ref{fig:all_calibration_data}.

\subsection{Phase noise calibration} \label{app:calibration}

Here we describe the phase noise calibration method (with the setup shown in Fig.~\ref{fig:setup}a) used to characterize the lasers. An electro-optic phase modulator is used to generate sidebands on the optical laser signal at $\omega_L$, by creating a phase modulation of 
\be
\phi(t) = \beta \cos(\omega_c t).
\ee
The ratio of the power between the carrier at $\omega_L$ and the first order sideband is given by 
\bea
\frac{P_1}{P_0}  = \left| \frac{J_1(\beta)}{J_0(\beta)} \right|^2.
\eea
Using a scanning Fabry-P\'erot filter with a bandwidth much smaller than $\omega_c$, we select out each sideband individually, and measure the powers $P_{0,1}$ in the carrier and sidebands to obtain a value for $\beta$.

The frequency noise spectrum for a known modulation $\phi(t)$ can be calculated from the Fourier transform of the autocorrelation function $\avg{\dot{\phi}(t)\dot{\phi}(t+\tau)}$.  For the case of sinusiodal phase modulation we have $\avg{\dot{\phi}(t)\dot{\phi}(t+\tau)}=\omega^2 \beta^2 \cos(\omega_c \tau) / 2$, with a corresponding frequency noise power spectral density of (in units of $\text{rad}^2\text{Hz}$),
\bea
\bar{S}^{\text{cal}}_{\omega\omega}(\omega) =\frac{\pi \omega^2 \beta^2}{2} \left( \delta(\omega - \omega_c) + \delta(\omega+\omega_c)\right).
\eea
By comparing the raw measured noise of the laser frequency noise ($S_\text{meas} (\omega)$) to that of the raw measured noise in the calibration signal peak ($S^{\text{cal}}_\text{meas} (\omega)$), one can calibrate the measured laser frequency noise in units of $\text{rad}^2\text{Hz}$.  Specifically, we have that $A(\omega_c) \int_{\omega_c - \Delta \omega}^{\omega_c + \Delta \omega} S^{\text{cal}}_\text{meas} (\omega) d\omega = \pi \omega^2 \beta^2/2$, where $A(\omega_c)$ is the conversion coefficient (at frequency $\omega_c$) between measured electrical noise power denity and (symmetrized) frequency noise power density for our experimental apparatus.   The corresponding laser frequency noise at $\omega_c$ can then be related to the measured electrical noise and the noise power in the phase modulation calibration tone as,
\bea
\bar{S}_{\omega \omega}(\omega_c) = \frac{\pi \omega_c^2 \beta^2}{2} \frac{ S_\text{meas} (\omega_c) }{\int_{\omega_c - \Delta \omega}^{\omega_c + \Delta \omega} S^{\text{cal}}_\text{meas} (\omega) d\omega}.
\eea
In order to generate the calibrated laser frequency noise spectra of Fig.~\ref{fig:all_data} we measured the conversion coefficient, $A(\omega_c)$, at $\sim 50$~MHz intervals across the entire frequency span of the measurement.    
\section*{References}
\bibliography{Mirror}

\end{document}